\documentclass[aps,pra,twocolumn,showpacs,groupedaddress,superscriptaddress,amsmath,amssymb,10pt,floatfix]{revtex4-1}
\usepackage{graphicx}% Include figure files
\usepackage{dcolumn}% Align table columns on decimal point
\usepackage{bm}% bold math
\usepackage{epsfig,amsmath}
\usepackage{subfigure}
\usepackage{color}
\usepackage{float}
\usepackage[normalem]{ulem}

\begin{document}
\title{Critical Transitions and Perturbation Growth Directions}

\author{Nahal~Sharafi}
\affiliation{Network Dynamics, Max Planck Institute for Dynamics and Self-Organization (MPIDS), 37077 G{\"o}ttingen, Germany}
\author{Marc ~Timme}
\affiliation{Network Dynamics, Max Planck Institute for Dynamics and Self-Organization (MPIDS), 37077 G{\"o}ttingen, Germany}
\affiliation{Technical University of Darmstadt, 64289 Darmstadt, Germany}
\affiliation{Institute for Nonlinear Dynamics, University of G{\"o}ttingen, 37077 G{\"o}ttingen, Germany}
\author{Sarah ~Hallerberg}	
\affiliation{Network Dynamics, Max Planck Institute for Dynamics and Self-Organization (MPIDS), 37077 G{\"o}ttingen, Germany}
\affiliation{Hamburg University of Applied Sciences, 20099 Hamburg, Germany}
%
%%%%%%%%%%%%%%%%%%%%%%%%%%%%%%%%%%%%%%%%%%%%%%%%%%%%%%%%%%%%%%%%%%%%%%%%%%%%%%%%%%%%%%%%
%

\begin{abstract}
Critical transitions occur in a variety of dynamical systems.
Here, we employ quantifiers of chaos to identify changes in the dynamical structure of complex systems preceding critical transitions.
As suitable indicator variables for critical transitions, we consider changes in growth rates and directions of covariant Lyapunov vectors.
Studying critical transitions in several models of fast-slow systems, i.e., a network of coupled FitzHugh-Nagumo oscillators, models for Josephson junctions and the Hindmarsh-Rose model, we find that tangencies between covariant Lyapunov vectors are a common and maybe generic feature during critical transitions.
We further demonstrate that this deviation from hyperbolic dynamics is linked to the occurrence of critical transitions by using it as an indicator variable and evaluating the prediction success through receiver operating characteristic curves.
In the presence of noise, we find the alignment of covariant Lyapunov vectors and changes in finite-time Lyapunov exponents to be more successful in announcing critical transitions than common indicator variables as, e.g., finite-time estimates of the variance.
Additionally, we propose a new method for estimating approximations of covariant Lyapunov vectors without knowledge of the future trajectory of the system. 
We find that these approximated covariant Lyapunov vectors can also be applied to predict critical transitions. 
\end{abstract}
\maketitle
%
%%%%%%%%%%%%%%%%%%%%%%%%%%%%%%%%%%%%%%%%%%%%%%%%%%%%%%%%%%%%%%%%%%%%%%%%%%%%%%%%%%%%%%%%
%
\section{Introduction}
Abrupt drastic shifts, called \textit{critical transitions} (CTs), have been reported in a variety of systems.~Seizures in epileptic patients 
\cite{Venegasetal,McSharrySmithTarassenko}, sudden crashes in financial markets \cite{MayLevinSugihara} and abrupt changes in climate \cite{Lentonetal,Alleyetal} and in ecosystems \cite{Scheffer} are all examples of critical transitions.~During a CT,
a system undergoes a sudden, relatively rapid and sometimes irreversible change.~A common model for systems exhibiting CTs are fast-slow systems \cite{KuehnBook}.
In this contribution we study CTs in several different fast-slow systems, such as the FitzHugh-Nagumo oscillator \cite{fitzhugh1961impulses},
a network of coupled FitzHugh-Nagumo oscillators, a model describing Josephson junctions \cite{jjp} and the Hindmarsh-Rose model \cite{hindmarsh1984model}.
Understanding a system exhibiting CTs as a dynamical system close to a bifurcation point, we can expect a CT to be preceded by \textit{early-warning signs} \cite{Wiesenfeld1,wissel1984universal}.~The most
famous symptom of CTs is the system's increasingly slow recovery from perturbations near the tipping point, a phenomenon referred to as \textit{critical slowing down} \cite{Wiesenfeld1,wissel1984universal,dakos2008slowing}.
Consequences of critical slowing down can be monitored in several observables that have been used to predict CTs.~Well studied indicator variabes are, e.g., \textit{increase in variance} \cite{Scheffer} and \textit{increase in autocorrelation} \cite{Scheffer} before a CT.
While these predictors are relatively successful in predicting CTs, they do not offer any insight into the dynamical structure of the phase space.
%
%%%%%%%%%%%%%%%%%%%%%%%%%%%%%%%%%%%%%%%%%%%%

In this contribution we investigate changes in the dynamical properties of different dynamical systems before a CT occurs.~We are especially interested in changes in the finite-time growth rates of perturbations and in changes in the directions of perturbation growth, as described by the stable and the unstable manifolds.
To explore these intrinsic directions of the phase space, we employ \textit{covariant Lyapunov vectors} \cite{Wolfe, Ginelli, Pazo, Parlitz} and \textit{finite-time Lyapunov exponents} \cite{pikovsky2016lyapunov}.
We investigate how changes in the finite-time Lyapunov exponents and changes in the intrinsic directions of the phase space, represented by covariant Lyapunov vectors, are linked to the occurrence of CTs in the aforementioned stochastic fast-slow models for critical transitions.
%
%%%%%%%%%%%%%%%%%%%%%%%%%%%%%%%%%%%%%%%%%%%%

As a characteristic feature, we find that tangencies between covariant Lyapunov vectors are linked to the occurrence of critical transitions.
The existence of tangencies (without any link to specific events) has been reported in models for spatio-temporal chaos as well \cite{HongLiuYang2009, Morris2013}.
Merging of covariant Lyapunov vectors indicates homoclinic tangencies between the stable and unstable manifolds which can occur in dynamical structures called \textit{wild hyperbolic sets} \cite{guckenbook}.
Newhouse \cite{newhouse1979abundance,gonchenko2002newhouse,yang2011newhouse} has proved the existence of hyperbolic invariant sets in which stable and unstable manifolds can have persistent homoclinic tangencies that are robust against perturbations.
A very recent contribution \cite{beims2016alignment} has reported tangencies to occur during transitions to distant branches of a trajectory for two three-dimensional deterministic models.
By evaluating the statistical relevance of the link between tangencies and CTs, we find the existence of these homoclinic tangencies to be a common phenomenon during critical transitions in two, three and higher-dimensional stochastic fast-slow models of critical transitions.
We argue that along with an increase in the first finite-time Lyapunov exponent, both observations can be expected to be intrinsically related to the phenomenon of critical slowing down.
We quantify this link in the sense of Granger causality by using changes in the directions of covariant Lyapunov vectors and their growth rates within prediction experiments.

Moreover, we develop and test a method for estimating approximations of covariant Lyapunov vectors without knowledge of the far future of the system which can be applied in predictive settings.
We find that properties of covariant Lyapunov vectors have the potential to be used as indicator variables for CTs since they perform equally well or, in the presence of noise, better than common indicator variables.
%
%%%%%%%%%%%%%%%%%%%%%%%%%%%%%%%%%%%%%%%%%%%%

This paper is organized as follows: In Sec.~\ref{clvs} we introduce covariant Lyapunov vectors and finite-time Lyapunov exponents and the methods we use for computing or approximating them.
In subsequent sections we present critical transitions and their footprints in covariant Lyapunov vectors and finite-time Lyapunov exponents for a single FitzHugh-Nagumo oscillator (Sec.~\ref{fhn}), 
the Hindmarsh-Rose model (Sec.~\ref{hmr}), a fast-slow model for Josephson junctions (Sec.~\ref{secjjj}) and a network of coupled FitzHugh-Nagumo oscillators 
(Sec.~\ref{fhnnet}). 
In Sec.~\ref{pred} we quantify the strength of links between the dynamics of covariant Lyapunov vectors and the occurence of critical transitions by evaluating the success of prediction experiments.
We summarize and discuss our findings in Sec.~\ref{conclusions}.
%
%%%%%%%%%%%%%%%%%%%%%%%%%%%%%%%%%%%%%%%%%%%%%%%%%%%%%%%%%%%%%%%%%%%%%%%%%%%%%%%%%%%%%%%%
%
\section{Covariant Lyapunov vectors} \label{clvs}
Covariant Lyapunov vectors are independent of the chosen coordinates, they are invariant under time reversal and covariant with the dynamics of the system.
They represent the stable and unstable manifolds and can provide information about the local structure of attractors \cite{Wolfe, Ginelli, Pazo, Parlitz}.
We compute covariant Lyapunov vectors for different fast-slow systems and use them to explore the dynamics.
Covariant Lyapunov vectors point in the directions of perturbation growth and live in the tangent space, the dimension of which is equal to the dimension of the original system.
The dynamics of the tangent space is governed by the linear propagator $F(t_1,t_2)$, which determines the evolution of perturbations $\delta u(t)$, i.e., $F(t_1,t_2)\, \delta u(t_1) = \delta u(t_2)$, see, e.g., \cite{Parlitz}.
Covariant Lyapunov vectors $\{\gamma_i(t)\}$ are the set of vectors whose evolution can be written in the form $ \|F(t_1,t_1\pm t) \gamma_i(t_1) \| \approx \mbox{exp}(\pm \mu_i(t))$ \cite{Parlitz}, $i = 1, ...m$, where $m$ is the dimension of the system under study and 
$\mu_{i}(t)$ denotes the instantaneous growth rate along the direction of the $i$-th covariant vector, the time-average of which is the $i$-th Lyapunov exponent, $\lambda_i$.
Covariant Lyapunov vectors are not only invariant under time reversal but also covariant with the flow.
Hence, in theory, once computed at one point, they can be determined at all times by $F(t_1,t_2) \gamma_i(t_1) = \gamma_i(t_2)$ \cite{Parlitz}.
In numerical computations the evolution of the vectors might be limited by the accumulation of numerical errors. 
In contrast to the orthogonal set of backwards Lyapunov vectors that are a byproduct of the process of computing Lyapunov exponents \cite{benettin1980lyapunov} covariant Lyapunov vectors are an orthorgonal set.
Consequently, one can study the dynamics of angles between covariant Lyapunov vectors and relate it to the dynamics of the system.
%
%%%%%%%%%%%%%%%%%%%%%%%%%%%%%%%%%%%%%%%%%%%%%%%%%%%%%%%%%%%%%%%%%%%%%%%%%%%%%%%%%%%%%%%%
%
\subsection{Computing covariant Lyapunov vectors} \label{compmeth}
Recently, several methods for computing covariant Lyapunov vectors have been proposed \cite{Ginelli,Parlitz,Wolfe}.
In this contribution we use the method introduced by Ginelli et al.~\cite{Ginelli} along with a new complementary method of approximating covariant Lyapunov vectors which will be introduced in Sec.~\ref{newmeth}.

The main idea of Ginelli et al.'s method is to perform iterations backwards in time on a random set of perturbation vectors confined to the subspaces spanned by backward Lyapunov vectors.
The idea stems from the fact that although forward and backward Lyapunov vectors \cite{benettin1980lyapunov,Parlitz} are not covariant with the dynamics, the subspaces they span, (Oseledec subspaces) are.
Assume $\phi_j^-(t)$ to be the $j$-th backward Lyapunov vector, growing asymptotically in time with exponential rate $\lambda_j$ and $j=1, ..., m$. Typically $\phi_j^-(t)$ is computed through Benetin's method \cite{benettin1980lyapunov} for computing Lyapunov exponents, $\lambda_i$.
Consider an arbitrary perturbation vector $\delta u_j(t_1) \in S^-_j(t_1) \backslash S^-_{j-1}(t_1)$ with Oseledec subspaces given by $S_j^-(t) = \text{span}\{\phi_i^-(t)\}, \thickspace i=0,...,j$.
Backward Lyapunov vectors can be computed by a QR decomposition after evolution with the linear propagator, i.e.,
\begin{eqnarray}
F(t_1,t_2)\bm{\Phi}^-(t_1) = \bm{\Phi}^-(t_2)\bm{R}(t_1,t_2),
\label{eqcv1}
\end{eqnarray}
with $\bm{\Phi}^-(t) = [\phi_1^-(t),\phi_2^-(t),\ldots,\phi_m^-(t)]$, $m$ being the dimension of the tangent space and $\bm{R}(t_1,t_2)$ is an upper triangular matrix.
On the other hand, we know that $\gamma_j(t)$, the $j$-th covariant Lyapunov vector, also belongs to the subspace $S^-_j(t) \backslash S^-_{j-1}(t)$. 
Hence, in matrix form, covariant Lyapunov vectors can be represented as,
\begin{eqnarray}
\bm{\Gamma}(t) = \bm{\Phi}^-(t)\bm{A}^-(t),
\label{eqcv2}
\end{eqnarray}
with $\bm{\Gamma}(t_1) = [\gamma_1(t),\gamma_2(t),\ldots,\gamma_m(t)]$  and $\bm{A}^-(t)$ being an upper triangular matrix.
Due to the covariance of the vectors, their evolution by the linear propagator can be described by
\begin{eqnarray}
F(t_1,t_2) \bm{\Gamma}(t_1) = \bm{\Gamma}(t_2)\bm{C}(t_1,t_2),
\label{eqcv3}
\end{eqnarray}
with $\bm{C}(t_1,t_2)$ denoting a diagonal matrix whose diagonal elements represent the growth rate of the vectors, i.e., covariant finite-time Lyapunov exponents.
Inserting Eq.~(\ref{eqcv2}) into Eq.~(\ref{eqcv3}) and using Eq.~(\ref{eqcv1}), we obtain,
\begin{eqnarray}
\bm{R}(t_1,t_2)\bm{A}^-(t_1) = \bm{A}^-(t_2)\bm{C}(t_1,t_2).
\label{eqcv4}
\end{eqnarray}
For backward iterations we have,
\begin{eqnarray}
\bm{R}(t_1,t_2)^{-1}\bm{A}^-(t_2) = \bm{A}^-(t_1)\bm{C}(t_1,t_2)^{-1}.
\label{eqcv5}
\end{eqnarray}
Eq.~(\ref{eqcv5}) enables us to perform iterations backwards in time while confining the iterations to the space of projections onto the backward vectors in order to converge to covariant Lyapunov vectors.

A practical implementation of this procedure is as follows:
Start iterating the system forward from the far past while orthogonalizing the perturbation vectors via QR decomposition every several time steps.~After the transient
time, the vectors converge to the backward Lyapunov vectors, $\bm{\phi}^-_i(t)$.
Start recording the backward vectors and $\bm{R}(t_1,t_2)$, the diagonal elements of which are the finite-time Lyapunov exponents.
Iterate to the far future.
Initialize a random nonsingular upper triangular matrix and iterate backwards with Eq.~(\ref{eqcv5}).
After the transient time, the upper triangular matrix will converge to $\bm{A}^-(t)$.
Use Eq.~(\ref{eqcv2}) to calculate covariant Lyapunov vectors.

\subsection{Estimating approximations of covariant Lyapunov vectors without knowledge of the (far) future}
\label{newmeth}
This contribution aims to quantify the existence of a link between tangencies of covariant Lyapunov vectors and critical transitions in a Granger causal sense, i.e. by testing in how far covariant Lyapunov vectors can predict critical transitions. 
However, since the computation of covariant Lyapunov vectors requires the knowledge of the future trajectory (see Sec.~\ref{compmeth}), these predictions would not be feasible in practical applications. 
Hence, we develope a new approach of estimating approximations of covariant Lyapunov vectors that does not require the knowledge of the (far) future of the system, but uses only data from the past and the near future.
The main idea of this approach is, to evolve covariant Lyapunov vectors computed at the preceding time step forward, in the space of projections onto backward Lyapunov vectors. 
Then evolve the resulting vectors from the near future (e.g. the next orthogonalization step) backwards repeatedly (without actually going backwards in time). 

Suppose that you would like to estimate an approximation of a covariant Lyapunov vector at time $t_n$ and you have data until time step $t_{n} + \tau$. 
The time $\tau$ can be as small as needed, the minimum value being one orthorgonalization step $\Delta$. 
Like other methods, to estimate covariant Lyapunov vectors we use two transient times, both however are in the past. 
The first transient is for the perturbation vectors to converge to the backward vectors and the second is for them to converge to the covariant vectors. 
We start in the far past and evolve the system and the perturbation vectors as explained before in Sec.~\ref{compmeth}. 
After a long enough transient time, one can assume that the perturbations have converged to backward Lyapunov vectors at time $t_n$. 
The next step is to compute the vectors that, through the second transient, will converge to covariant Lyapunov vectors. 
The second transient does not need to be long, it can be as short as the available data allows. 
Knowing this dynamics one can start the second transient at time $t_n$, i.e., evolve the perturbation vectors from $t_n$ to $t_n + \tau$ and record $\bm{R}(t_n,t_n+\tau)$. 
%
%The time difference $\tau$ can be as small as needed, the minimum value being $\Delta$. i.e. one orthogonalization step. 
%
Eq.(\ref{eqcv4}) and Eq.(\ref{eqcv5}) determine dynamical rules of the backwards and forward transformation of the covariant Lyapunov vectors in the space of projections onto backward Lyapunov vectors.
Knowing this dynamics one can continue evolving the perturbation vectors from $t_n$ to $t_n + \tau$ and record $\bm{R}(t_n,t_n+\tau)$. 
We then compute the inverse, $\bm{R}(t_n,t_n+\tau)^{-1}$ and multiply it repeatedly with a random upper-triangular matrix $\bm{A_R}(t_n)$ according to Eq.~(\ref{eqcv5}).
In other words, we obtain the first approximated matrix $\bm{A}^{'}(t_n)$ through
\begin{eqnarray}
\bm{A}^{'}(t_n) \propto \bm{R}(t_n,t_n+\tau)^{-N} \bm{A_R}(t_n)    
\label{eq68}
\end{eqnarray}
with $N$ being adjusted with respect to the model we are studying. 
Multiplying $A^{'}(t_n)$ with the matrix of backward vectors gives approximations of covariant Lyapunov vectors at time $t_n$.

After evaluating the vectors for the very first time step of the second transient, $t_n$, one does not have to use the evolution of random matrices or matrices of eigenvectors any more. 
As a matter of fact one can improve the estimate of the covariant vectors by evolving the vectors computed at the previous time step and then use the evolved matrix as a starting matrix for going backwards. 
In more detail, knowing $\bm{A}^{'}(t_n)$, the matrix $\bm{A}^{'}(t_{n+1})$ for the next time step $t_{n+1} = t_n + \Delta $, can be obtained by evolving $\bm{A}^{'}(t_n)$ from $t_n$ to $t_{n+1}+\tau$,
in the space of projections onto the backward vectors using Eq.~(\ref{eqcv4}), and then using the evolved matrix to iterate backwards from $t_{n+1}+\tau$ to $t_{n+1}$, $N$ times. Since 
$\bm{R}(t_n,t_{n+1})^{-1}\bm{R}(t_n,t_{n+1}) = 1$ we have,
\begin{eqnarray}
\bm{A}^{'}(t_{n+1}) \propto \bm{R}(t_{n+1},t_{n+1}+\tau)^{-(N-1)}\bm{R}(t_n,t_{n+1}) \bm{A^{'}}(t_n), \quad  
\label{eq70}
\end{eqnarray}
Using information of the vectors computed in the past one can therefore improve the estimate of the vectors in the present. 
Therefore using matrices $A^{'}(t_p)$ from the past to successively compute $A^{'}(t_n)$ with $t_p < t_n$, gradually improves the precision of the approximative estimates during the time steps of the second transient .
After this short second transient one can assume that the matrix $\bm{A}^{'}(t)$ has converged to an estimate of $\bm{A}^-(t)$, that is the matrix of projection of covariant Lyapunov vectors onto the backward vectors. 
\begin{figure}[]   
 \centerline{
\includegraphics[angle=0,width=0.4\textwidth]{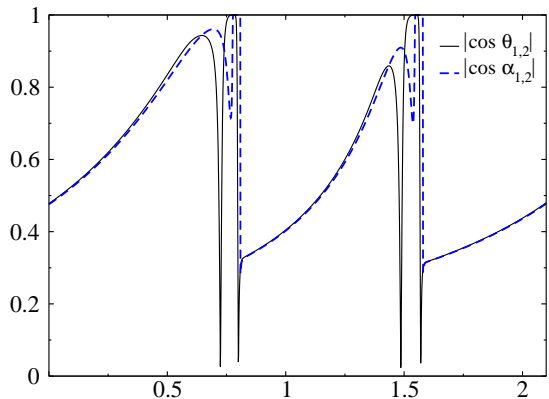}
}
\caption{ 
 Estimating approximations of covariant Lyapunov vectors with the repetitive iteration method yields results which mimic the dynamics of covariant Lyapunov vectors computed through Ginelli et. al.'s method.
The absolute value of the cosine of the angle between the first and the second covariant Lyapunov vectors computed with Ginelli et al.'s method, solid black line, are compared with the results from the repetitive iteration method, dashed blue line, for a FitzHugh-Nagumo model with $a = 0.4, b=0.3, e = 0.01$ and $D = 0$.
}
\label{figs:samfhncomp}
\end{figure}
A difficulty that can occur with this method in systems with highly expanding directions is that $\bm{A}^{'}$ becomes ill-conditioned and several vectors collapse on each other. 
The solution in that case is to randomize the matrix $\bm{A}^{'}$ again. 
Ill-conditioned $\bm{A}^{'}$ can also occur during repeated iterations. 
This problem usually occurs if $N$ is too large. 
In this case it suffices to reduce $N$ to a value that is large enough to guarantee convergence of the vectors to the covariant directions and small enough not to cause singularities in $\bm{A}^{'}$. 
Note that if one has enough data in the past one can use any method to compute covariant vectors (e.g., Ginelli et al.'s) until the near past and then use the approximative method described above to compute a present estimate of covariant vectors. 

Moreover bear in mind that this method only yields approximations of covariant Lyapunov vectors and not the exact vectors. 
Nonetheless it seems to be leading to effective approximations concerning the dynamics of the vectors. 
Figure \ref{figs:samfhncomp} shows the absolute value of the cosine of the angle between first and second covariant Lyapunov vector $\theta_{1,2}$ for a single FitzHugh-Nagumo oscillator (see Sec.~\ref{eqfhn} for details) computed with Ginelli et~al.'s method as well as the corresponding angle $\alpha_{1,2}$ obtained from the approximative method presented in this section.
In Sec.~\ref{pred} we will compare the predictions made using time series of the  angle $\theta_{ij} (t)$ between the covariant Lyapunov vectors (computed as described in Sec.~\ref{compmeth}) to predictions based on time series of the angle $\alpha_{ij} (t)$ between approximations of covariant Lyapunov vectors as described in this section. 
%%%%%
%%%%%%%%%%%%%%%%%%%%%%%%%%%%%%%%%%%%%%%%%%%%%%%%%%%%%%%%%%
\subsection{Computing finite-time Lyapunov exponents}\label{ftles}
Lyapunov exponents $\lambda_i$ are well known quantifiers of chaos that measure the average growth rate of perturbations in different directions.
While Lyapunov exponents are time-averaged quantities, their finite-time values $\{\lambda_i(t)\}$ , i.e., finite-time Lyapunov exponents, can describe the current behavior of the system under study \cite{shadden2005definition,arnold1986lyapunov}.
Although Lyapunov exponents $\lambda_i$, computed in the asymptotic limit of infinite-time are well ordered by value, their finite-time counterparts, the finite-time Lyapunov exponents $\lambda_i(t)$, can fluctuate and exchange order.
In addition to covariant Lyapunov vectors and their growth rates $\{\mu_i(t)\}$, we compute finite-time Lyapunov exponents $\{\lambda_i(t)\}$ \cite{pikovsky2016lyapunov} which represent growth rates of backward Lyapunov vectors.
For computing these exponents in the low-dimensional systems studied in this contribution we used the Euclidean norm.
Since finite-time Lyapunov exponents describe the current behavior of the system under study \cite{shadden2005definition,arnold1986lyapunov}, we will study here whether changes in these exponents can be signatures announcing critical transitions.
General considerations on Lyapunov exponents in oscillators with noise can e.~g.~ be found in \cite{Goldobin2005} and in \cite{pikovsky2016lyapunov}.

Concerning further technical details of all computations: for all the systems studied in this contribution, we integrated the equations using a forth-order Runge-Kutta solver.
Our integration time step was $dt = 0.001$ and we orthogonalized the perturbation vectors after every 10 iteration steps $\delta t = 0.01$.
When using the newly developed approach for estimating approximations of covariant Lyapunov vectors, we iterated the system forward in time only one orthogonalization interval i.e. $\tau = \Delta$.
% 
% 
%%%%%%%%%%%%%%%%%%%%%%%%%%%%%%%%%%%%%%%%%%%%%%%%%%%%%%%%%%%%%%%%%%%%%%%%%
%%%%%%%%%%%%%%%%%%%%%%%%%%%%%%%%%%%%%%%%%%%%%%%%%%%%%%%%%%%%%%%%%%%%%%%%%
\section{Critical Transitions in a FitzHugh-Nagumo Oscillator}
\label{fhn}
%%%%%%%%%%%%%%%%%%%%%%%%%%%%%%%%%%%%%%%%%%%%%%%%%%%%%
\begin{figure*}[ht!!!]
% %
% \begin{minipage}[t]{0.7\textwidth}
\centerline{
\includegraphics[width=1\textwidth]{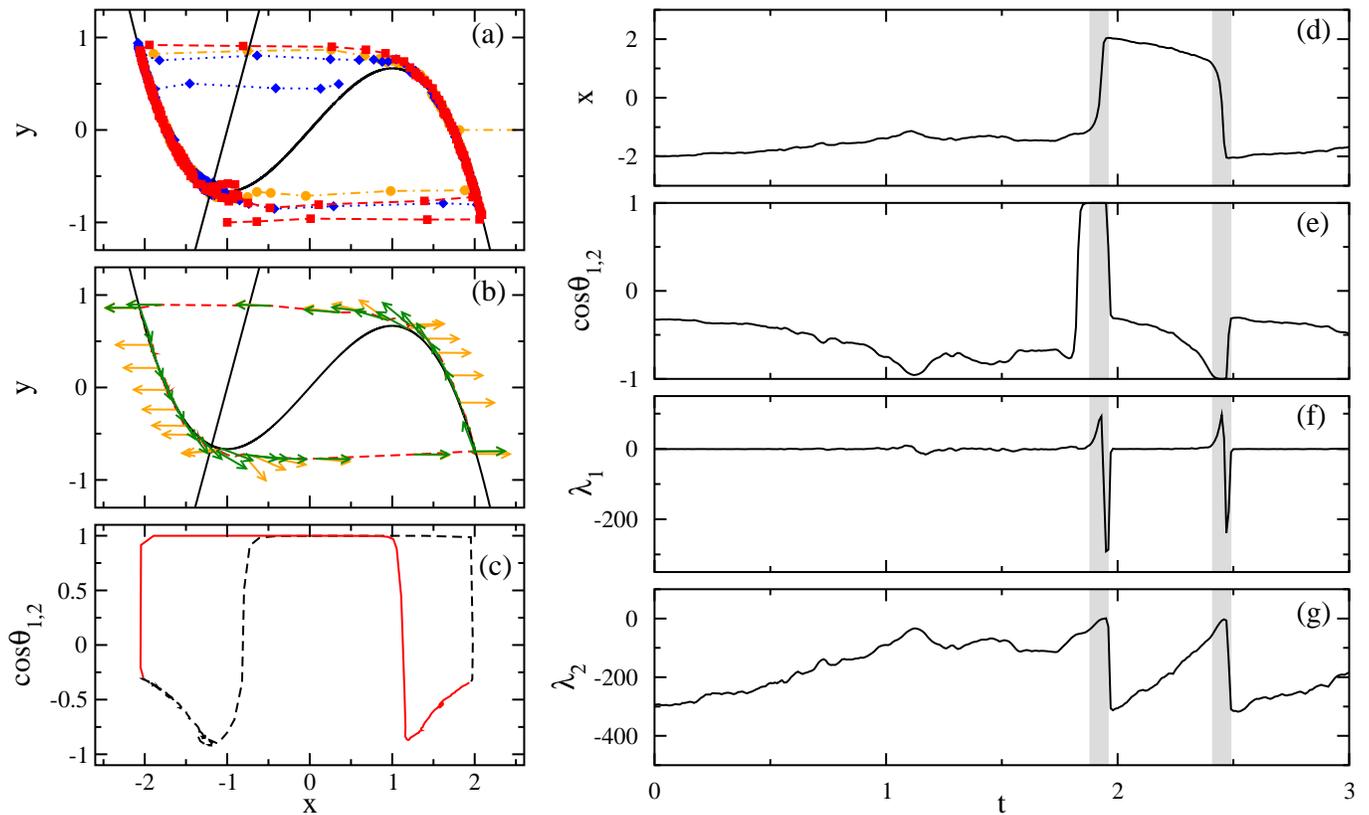}
}%, height=0.9\textwidth
%
% \end{minipage}
\caption{\label{fig:fhn02dni} Alignment of covariant Lyapunov vectors during noise-induced transitions in a FitzHugh-Nagumo oscillator with $a = 1$, $b = 0.3$, $\epsilon = 0.01$ and $D = 0.2$.
(a) Phase space portrait of the oscillator with black lines indicating the nullclines and the discontinuous lines representing trajectories from different initial points.
(b) The dotted red line indicates a typical trajectory in the phase space.
Green and orange vectors show the first and the second covariant Lyapunov vector respectively, both vectors align during transitions.
(c) The cosine of the angle between the first and the second vector is shown while the system is drifting on the left and transitioning to the right branch (black dashed line) and drifting on the right and transitioning to the left branch (red line).
(d) Time series of the fast variable, i.e., the observable of the system.
(e) Time series of the cosine of the angle between the first and the second vector.
(f) Time series of the first finite-time Lyapunov exponent and (g) Time series of the second finite-time Lyapunov exponent.}
\end{figure*}
%
%
%%%%%%%%%%%%%%%%%%%%%%%%%%%%%%%%
%
%
Fast-slow dynamical systems are common models for describing CTs \cite{KuehnBook}.
They show a slow drift of system variables interrupted by fast transitions, i.e., drastic changes in at least one variable.
The first fast-slow system we investigate is a FitzHugh-Nagumo oscillator \cite{fitzhugh1961impulses},
\begin{eqnarray}
\epsilon\dot{x} & = & x - \frac{x^3}{3}-y,\\
\dot{y} & = & x + a -by + \sqrt{2D}\eta(t), \label{eqfhn}
\end{eqnarray}
where $\epsilon \ll 1 $. 
The slow variable $y$ acts as a bifurcation parameter and drives the dynamics of the fast variable $x$ such that it alternates between two different states with $\eta(t)$ being white noise.
We investigated this model for two different parameter sets.
The first set of parameters is $a = 1$, $b = 0.3$ and $\epsilon = 0.01$.
For this set, transitions can only be induced by noise.
%
% We chose the noise strength to be $D = 0.2$.
%
An example for a transition can be seen in the time series of the fast variable as presented in Fig.~\ref{fig:fhn02dni}(d).
In Fig.~\ref{fig:fhn02dni}(a) we present trajectories of a single oscillator in phase space with different initial conditions.
The black lines are nullclines, the polynomial curve is the nullcline of the fast variable, $x$, and the straight line is the nullcline of the slow variable, $y$.
For this set of parameters, the intersection of the nullclines is a stable fixed point to which trajectories would converge in the absence of noise.
However, the stochastic term added to the control parameter of Eq.~(\ref{eqfhn}) enables transitions.

Fig.~\ref{fig:fhn02dni}(b) shows the phase space portrait of the oscillator.
The dashed red line indicates the trajectory.
The green and the orange vectors are the first and the second covariant Lyapunov vectors respectively.
The first vector is tangent to the trajectory and represents the ”neutral” direction corresponding to the first Lyapunov exponent.
The second vector indicates the stable direction, perturbations along which decay rapidly.

In the following, we will focus for the moment on the behavior of the finite-time Lyapunov exponents.
While the system moves along the nullcline, the first finite-time Lyapunov exponent is close to zero before a transition [see Fig.~\ref{fig:fhn02dni}(f)].
Perturbations along the trajectory neither shrink nor grow.
Before the occurrence of a transition, the direction orthogonal to the trajectory is stable, consequently, the second finite-time Lyapunov exponent is negative [see Fig.~\ref{fig:fhn02dni}(g)].
As the system moves towards the transition point on the nullcline, the second finite-time Lyapunov exponent grows since the shrinking of the perturbation along the stable direction becomes slower as the system approaches the critical point.
During the transition, any perturbation orthogonal to the trajectory neither shrinks nor grows.
Hence the second finite-time Lyapunov exponent approaches zero [see Fig.~\ref{fig:fhn02dni}(g)].
Very close to the transition, the first finite-time Lyapunov exponent rapidly increases and becomes positive [see Fig.~\ref{fig:fhn02dni}(f)].
The phenomenon becomes apparent inspecting the x-nullcline in Fig.~\ref{fig:fhn02dni}(b): as the fast variable accelerates towards the other branch of the manifold, any perturbation along the trajectory will also rapidly grow.
However, during the second half of the transition, any perturbation along the trajectory will shrink rapidly as the fast variable is decelerating prior to arriving at the other branch.
At the end of the transition, arriving again at the x-nullcline, the first finite-time Lyapunov exponent approaches zero once more.

The transition is also reflected in the angle between first and second covariant Lyapunov vector, $\theta_{12}$.
% %
The angle decreases as the system moves towards a transition on the nullcline since the angle between the trajectory and the unstable direction is decreasing [see Fig.~\ref{fig:fhn02dni}(b) and Fig.~\ref{fig:fhn02dni}(c)].
% %
As the transition starts, the trajectory becomes completely tangent to the shrinking direction, moving rapidly towards the other branch of the nullcline.
% %
Consequently, the two covariant Lyapunov vectors are almost tangent [see Fig.~\ref{fig:fhn02dni}(b)].
% %
Note that this implies that the system is effectively one-dimensional during the transition to the other section of the manifold.
The $j$-th covariant Lyapunov vector as discussed in section \ref{newmeth} and \ref{compmeth} is the vector that grows backwards and forward in time with growth rate $\mu_j(t)$.
In the limit of infinite-time, these growth rates converge to the Lyapunov exponent $\lambda_j$.
Going backwards in time, the $j$-th covariant Lyapunov vector can be regarded as a vector belonging to the subspace $S^-_j(t)\backslash S^-_{j-1}(t)$. 
Therefore it is a linear combination of the first $j$ backward vectors with a nonzero component along the $j$ the backward vector and it is orthogonal to backward vectors of order higher that $j$ (see eq.\ref{eqcv2}).  
Therefore going backwards in time it asymptotically decays with the rate $\lambda_j$, i.e the smallest Lyapunov exponent of the exponents $\lambda_i$, $i=1 \dots j$. 
Although Lyapunov exponents $\lambda_i$, computed in the asymptotic limit of infinite-time are well ordered by value, their finite-time counterparts, the finite-time Lyapunov exponents $\lambda_i(t)$, can fluctuate and exchange order.
The direction corresponding to $\lambda_i(t)$ (the $i$-th backward Lyapunov vector $\phi^-_i(t)$) where $i < j$, may temporarily become more stable than the direction corresponding to $\lambda_j(t)$, i.e. the value of $\lambda_i(t)$ may temporarily be smaller than $\lambda_j(t)$.

In case for any reason the order between finite-time Lyapunov exponents is temporarily lost,  
any covariant vector of the order between $i$ and $j$, will have a dominant component along the $i$-th backward vector and tend to converge to the subspace $S^-_i(t)\backslash S^-_{i-1}(t)$,
forming tangencies with the $i$-th covariant vector.
In this contribution we argue that this temporary change in the stability of stable and unstable or neutral directions is a generic behavior in critical transitions that leads to tangencies between stable and unstable (or marginal) manifolds.

In the case of the FitzHugh-Nagumo, as the system slowly moves towards a transition point, the increase in the second finite-time Lyapunov exponent goes along with a decrease of the angle between the first and the second covariant Lyapunov vector. 
Right before and at the very beginning of the transition, the marginal direction becomes highly unstable, enabling the transition. 
The sudden rise of the first Lyapunov exponent making this exponent much larger than the second finite-time Lyapunov exponent which is close to zero, goes along with a fast increase in the angle between the two vectors. 
However this sudden rise is followed by a sharp decrease way below the value of the second finite-time Lyapunov exponent during the transition. 
The first Lyapunov exponent becoming the more negative exponent, is like a switching between the stability of the orthogonal directions of the finite-time Lyapunov exponents. 
The direction of the first finite-time Lyapunov exponent has temporarily become more stable 
than the direction orthogonal to it. 
Therefore the second covariant vector will collapse on the direction parallel to the previously marginal manifold.

\begin{figure}[t]
\centerline{
\includegraphics[angle=0,width=0.5\textwidth]{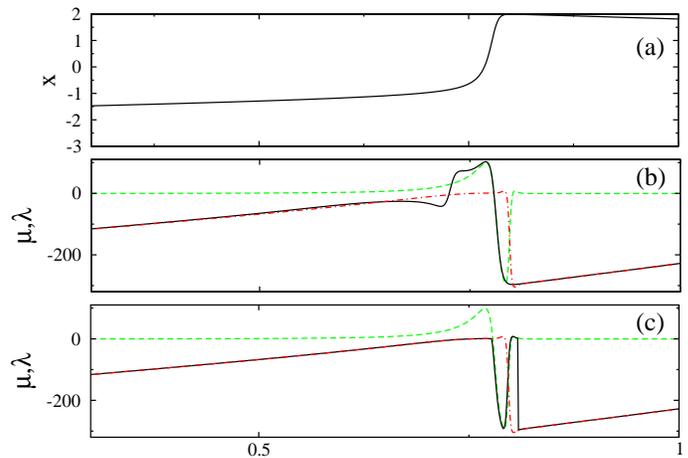}
}
\caption{\label{figs:fhn02dniclext}
The finite-time growth rate of the first and second covariant Lyapunov vector converge during the transition in a FitzHugh-Nagumo oscillator.
The first (dotted green line) and the second (dotted red line) finite time Lyapunov exponent are compared to the finite-time growth rate of the second covariant Lyapunov vector (solid black line). 
Note that the growth rate of the first covariant Lypunov vector is by definition the same as the first finite-time Lyapunov exponent. 
While the trajectory is slowly drifting on the nullcline the growth rate of the second covariant vector is almost the same as the second finite-time Lyapunov exponent. 
Merging of the first and the second covariant Lyapunov vector during the transition manifests it's self in converging of the growth rate of the second covariant vector to the first finite-time exponent.
(a) The time series of the fast
variable.
(b) The growth rates computed with Ginelli et. al.’s method
(c)The growth rates computed with repetitive iterations method. 
}
\end{figure}

Fig.~\ref{figs:fhn02dniclext} shows the finite-time exponents and the growth rate of the second covariant vector computed with Ginelli et. al.'s method and the repetitive iteration method.
Note that in order to be able to compare the results for the exact same transition, i.e $D = 0$, we simulated the FitzHugh-Nagumo model with $a = 0.4$ that corresponds to the regular spiking regime. 
The results with both methods show, as discussed before that the growth rate of the second vector is close to the second finite-time Lyapunov exponents while the trajectory is on one of the branches. 
During the transition however the growth rate of the second covariant Lyapunov vector converges to the first Lyapunov exponent. 
With the repetitive iteration method this convergence corresponds exactly to the instance that the first finite-time Lyapunov exponent becomes smaller than the second finite-time Lyapunov exponent,
which agrees with our discussion above. 
As for the Ginelli et. al.'s method they seem to converge earlier. The reason for that is in the repetitive iteration method, repeated iteration of the same interval at 
present amplifies the changes in the local dynamics, while in Ginelli et. al.'s method iterating backwards from the far future will lead to a delay in exhibiting the changes in the local dynamics.

%
%%%%%%%%%%%%%%%%%%%%%%%%
\begin{figure*}[ht]
% %
% \begin{minipage}[t]{0.7\textwidth}
\centerline{
\includegraphics[width=1\textwidth]{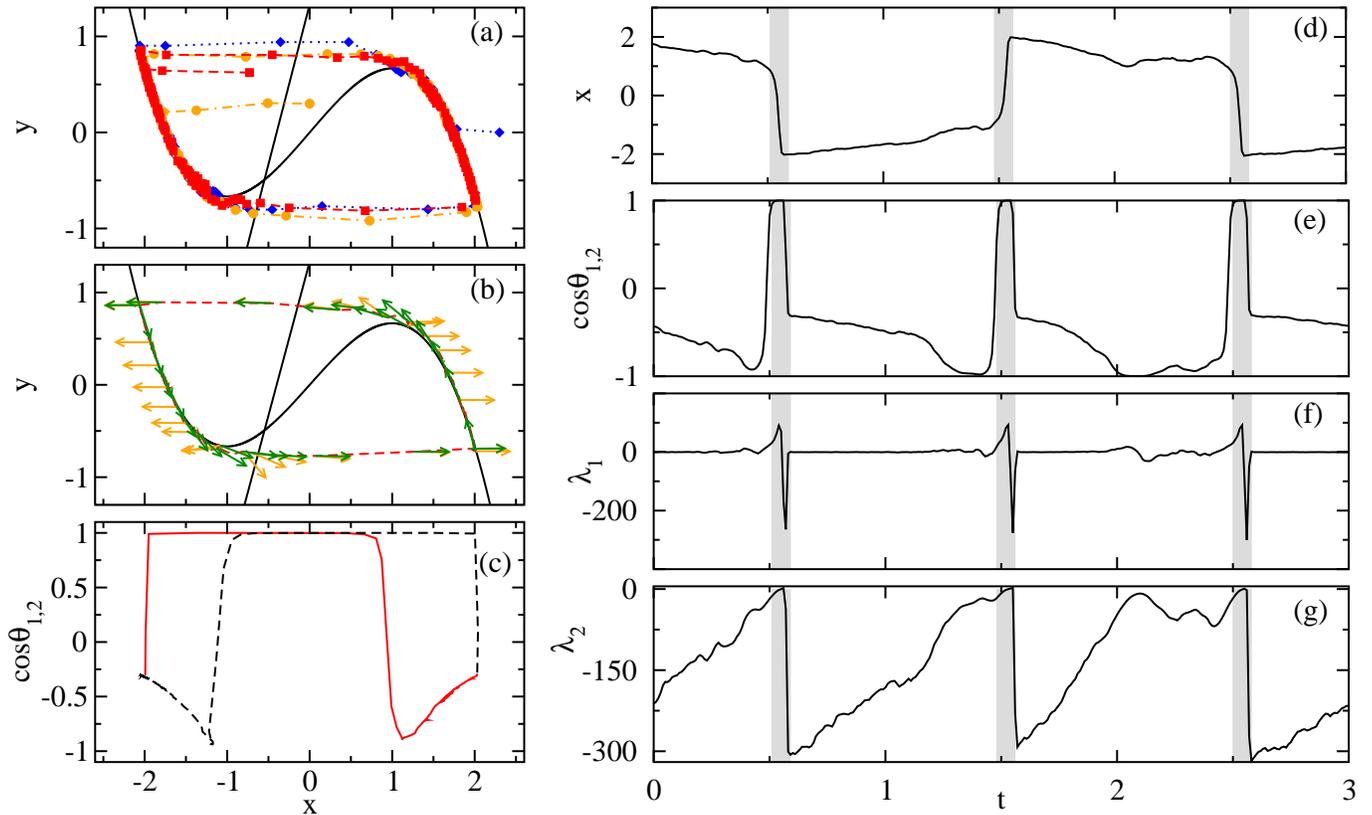}
}%, height=0.9\textwidth
%
% \end{minipage}
\caption{\label{fig:fhn02d}
Alignment of covariant Lyapunov vectors in a single FitzHugh-Nagumo oscillator with $a = 0.4$, $b = 0.3$, $\epsilon = 0.01$ and $D = 0.2$.
(a) Phase space portrait of the oscillator with black lines indicating the nullclines and the discontinuous lines representing trajectories from different initial points.
(b) The dotted red line indicates a typical trajectory in the phase space.
The green and the orange vectors show the first and the second covariant Lyapunov vector respectively, both vectors align during transitions.
(c) The cosine of the angle between the first and the second vector is shown while the system is drifting on the left and transitioning to the right branch (black dashed line) and drifting on the right and transitioning to the left branch (red line).
(d) Time series of the fast variable, i.e., the observable of the system.
(e) Time series of the cosine of the angle between the first and the second vector.
(f) Time series of the first finite-time Lyapunov exponent and (g) Time series of the second finite-time Lyapunov exponent.}
\end{figure*}
%%%%%%%%%%%%%%%%%%%%%%
%
We obtain qualitatively similar results concerning the dynamics of the angle between the covariant Lyapunov vectors and finite-time Lyapunov exponents for different sets of parameters, as, e.g., $a = 0.4, b = 0.3$ and $\epsilon = 0.01$, for which the fixed point is unstable and oscillations are present even in the absence of noise [see Fig.~\ref{fig:fhn02d}].
%%%%%%%%%%%%%%%%%%%%%%%%%%%%%%%%%%%%%%%%%%%%%%%%%%%%%%%%%%%%%%%%%%%%
% \clearpage
%%%%%%%%%%%%%%%%%%%%%%%%%%%%%%%%%%%%%
\section{Critical Transitions in the Hindmarsh-Rose Model}
\label{hmr}
%
%%%%%%%%%%%%%%%
\begin{figure}[ht]
\begin{subfigure}{}
\includegraphics[width=0.98\linewidth]{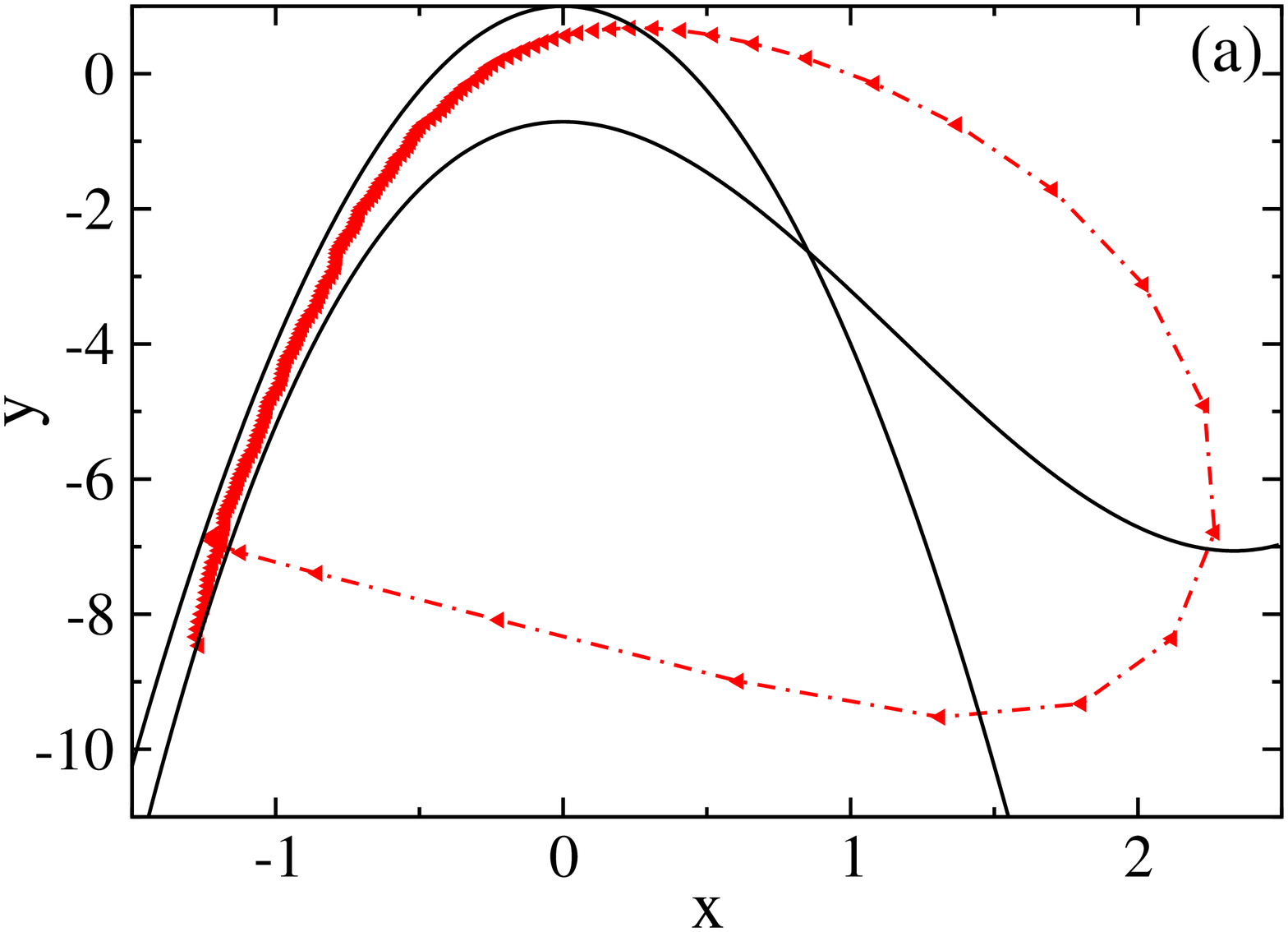}
\end{subfigure}
\begin{subfigure}{}
\includegraphics[width=0.98\linewidth]{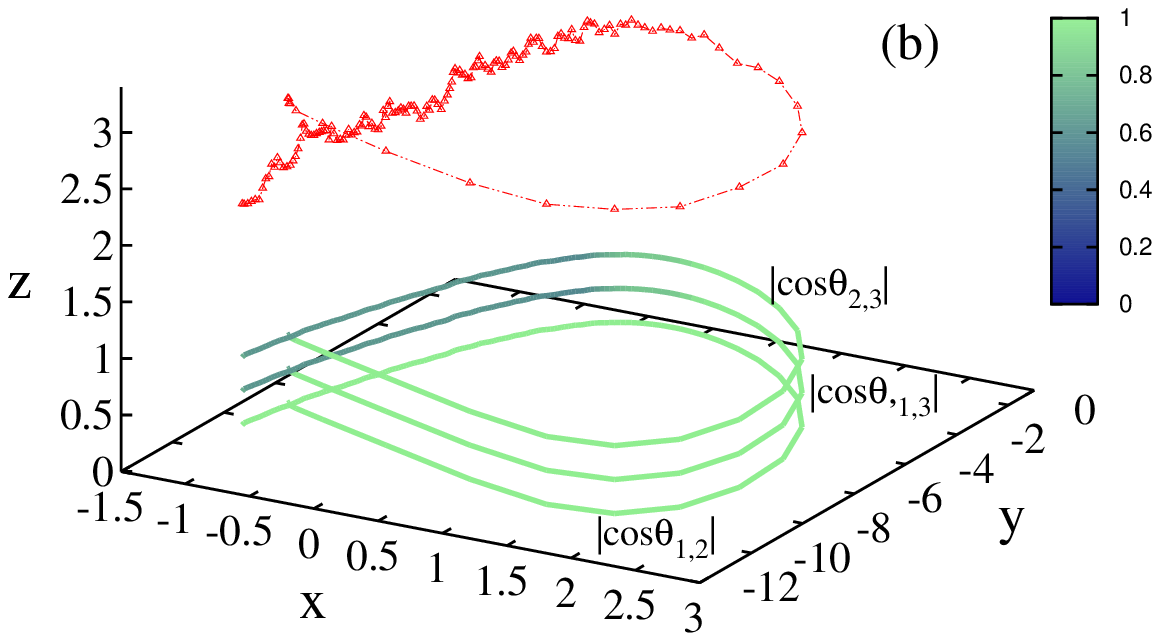}
\end{subfigure}
\caption{Merging of all covariant Lyapunov vectors forecasts spikes in the Hindmarsh-Rose model.
(a) Projection of the trajectory of the Hindmarsh-Rose model and the nullclines of $x$ and $y$ on the $x$-$y$-plane.
The trajectory (dash-dotted line) slowly moves up close to the nullclines (black solid lines) before the spike occurs.
(b) The three-dimensional phase space portrait of the trajectory shows a slow drift of $z$ as the fast variables go through an excursion.
The green-blue lines are the shifted projections of the trajectory on the $x$-$y$-plane, showing the absolute value of the cosine of the angle between the covariant Lyapunov vectors.}
\label{fig:hrphase}
\end{figure}
%%%%%%%%%%%%%%%%%%%%%%%%%%%%%%%%%%%%%%%%%%%%%%%%%%%%%%%%%%%%%%%%%
%%%%%%%%%%%%%%%
\begin{figure}[ht]
\begin{subfigure}{}
\includegraphics[width=0.98\linewidth]{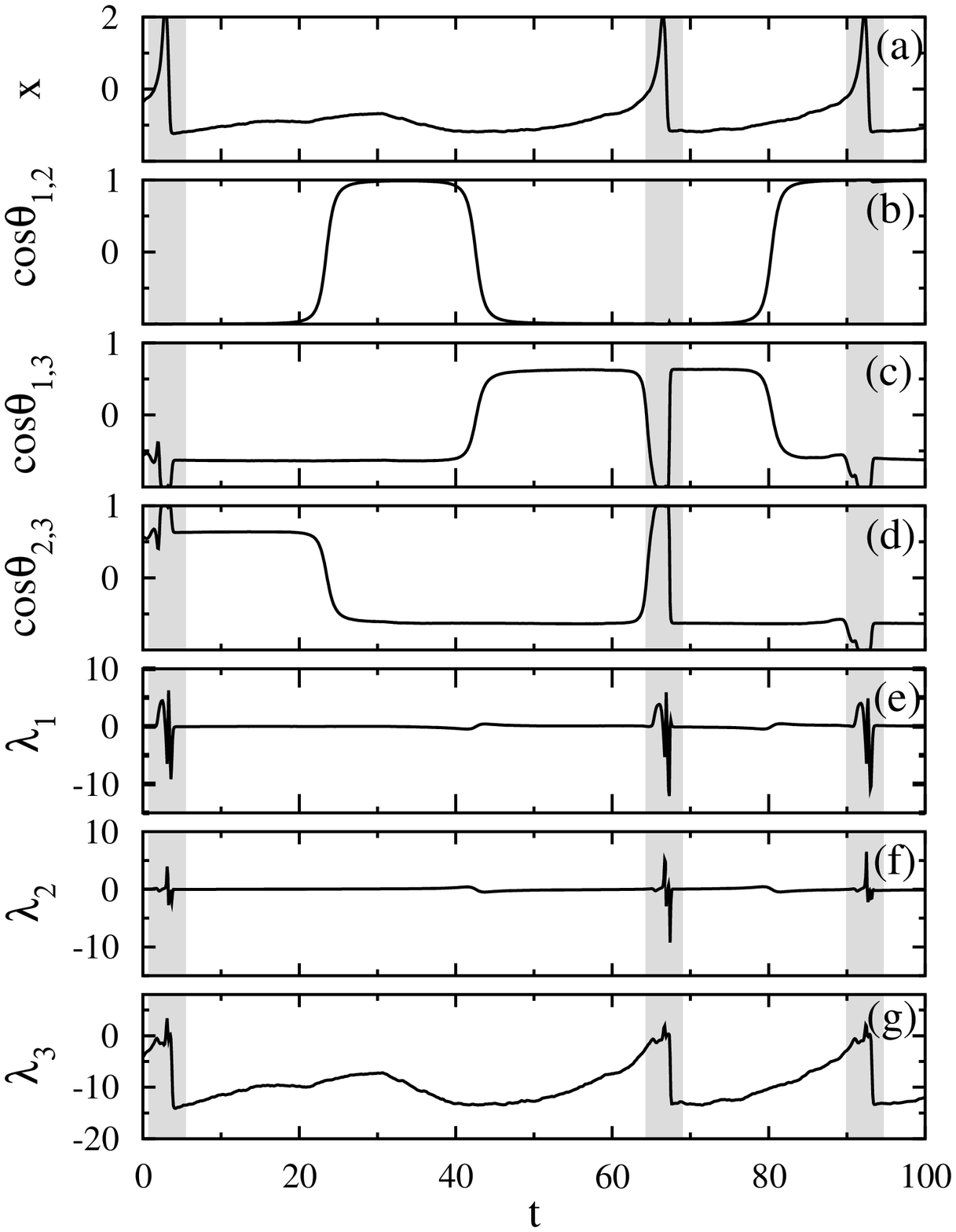}
\end{subfigure}
\caption{In the Hindmarsh-Rose model, all three covariant Lyapunov vectors merge during a critical transition and the finite-time Lyapunov exponents increase prior and reach their maximum during transitions.
(a) Time series of the fast variable $x$.
(b-d) Time series of the cosine of the angles between the covariant Lyapunov vectors.
(e-g) Time series of the finite-time Lyapunov exponents.}
\label{fig:hr}
\end{figure}
%%%%%%%%%%%%%%%%%%%%%%%%%%%%%%%%%%%%%%%%%%%%%%%%%%%%%%%%%%%%%%%%%
In order to test whether the alignment of covariant Lyapunov vectors is inherent to models that show CTs, we computed covariant Lyapunov vectors and finite-time Lyapunov exponents for trajectories of the Hindmarsh-Rose model \cite{hindmarsh1984model}.
%.
The Hindmarsh-Rose model is a model common to describe neural activity.
A modified version of this model, containing an additional stochastic term, is given by the following equations,
\begin{eqnarray}
\dot{x} & = & y - ax^3 + bx^2 -z + J, \\
\dot{y} & = & c - dx^2 - y, \\
\dot{z} & = & r(s(x-x_0)-z) + \sqrt{2D}\eta(t), 
\label{eq:hr}
\end{eqnarray}
where $x$ is a voltage-like variable, $y$ controls the recovery after a spike and $z$ represents an adapting current with slow dynamics.
We choose a noise strength of $D = 0.05$ and parameter values, $r = 0.01, s = 4, x_0 = -1.6, b = 3.5, a = 1, c = 1, J = 2.5$ and $d = 5$.
Within this parameter range, the original Hindmarsh-Rose model shows a regular spiking behavior.
However, our modification of the model, which consists of adding a stochastic term to the slow variable, leads to highly irregular spiking.

Fig.~\ref{fig:hrphase}(a) illustrates projections of the nullclines of $x$ and $y$ and the trajectory to the $x$-$y$-plane.
While the trajectory slowly moves close to the nullclines, the drifting of the bifurcation parameter $z$ shifts the $x$-nullcline and enables an excursion of $x$ and $y$, called a spike [also see Fig.~\ref{fig:hrphase}(b)].

Fig.~\ref{fig:hr} shows time series of the observable $x$, of all angles between the covariant Lyapunov vectors and of all finite-time Lyapunov exponents computed for the Hindmarsh-Rose model.
The first covariant Lyapunov vector, corresponding to the first Lyapunov exponent that is zero [see Fig.~\ref{fig:hr}(e)], is tangent to the trajectory.
However, for the Hindmarsh-Rose model, the second Lyapunov exponent [see Fig.~\ref{fig:hr}(f)] is also very small (-0.041 for this parameter set), and its finite-time value is also mostly close to zero and 
coinciding with the first finite-time exponent, resulting the persistent tangency between the first and the second covariant Lyapunov vector [see Fig.~\ref{fig:hr}(b)].
The 180 degrees change of the direction of the first and second vector is due to an intersection between the two finite-time exponents, i.e the switching of the stability of the corresponding backward Lyapunov vectors, that can be seen in 
Fig.~\ref{fig:hr}(e) and (f). However it is an insignificant change in the stability that does not correspond to the critical transition.

In this system the vector corresponding to the contracting direction is the third covariant Lyapunov vector.
The third finite-time exponent is not highly negative either, there for the third covariant Lyapunov vector has noticeable components along the first and the second covariant vectors.
Nonetheless the spikes or the critical transitions are marked by clear tangencies between the first and the third and the second and the third covariant Lyapunov vectors.

That is to say, the phenomenon of critical slowing down prior to CTs manifests itself in tangencies between the third and the first two covariant Lyapunov vectors during CTs [see Figs.~\ref{fig:hr}(b-d)].

The increase in finite-time Lyapunov exponents prior to CTs and the spike-like dynamics during the CTs, which have been observed for the FitzHugh-Nagumo oscillator studied in the previous section, also occur in time series of finite-time Lyapunov exponents computed from the Hindmarsh-Rose model.
% % %
%%%%%%%%%%%%%%%%%%%%%%%%%%%%%%%%
\section{Josephson Junctions} \label{secjjj}
\begin{figure*}[ht]
%
% \hspace*{0.13cm}
%
\begin{minipage}{0.48\textwidth}
\centerline{
\includegraphics[width=1.0\textwidth]{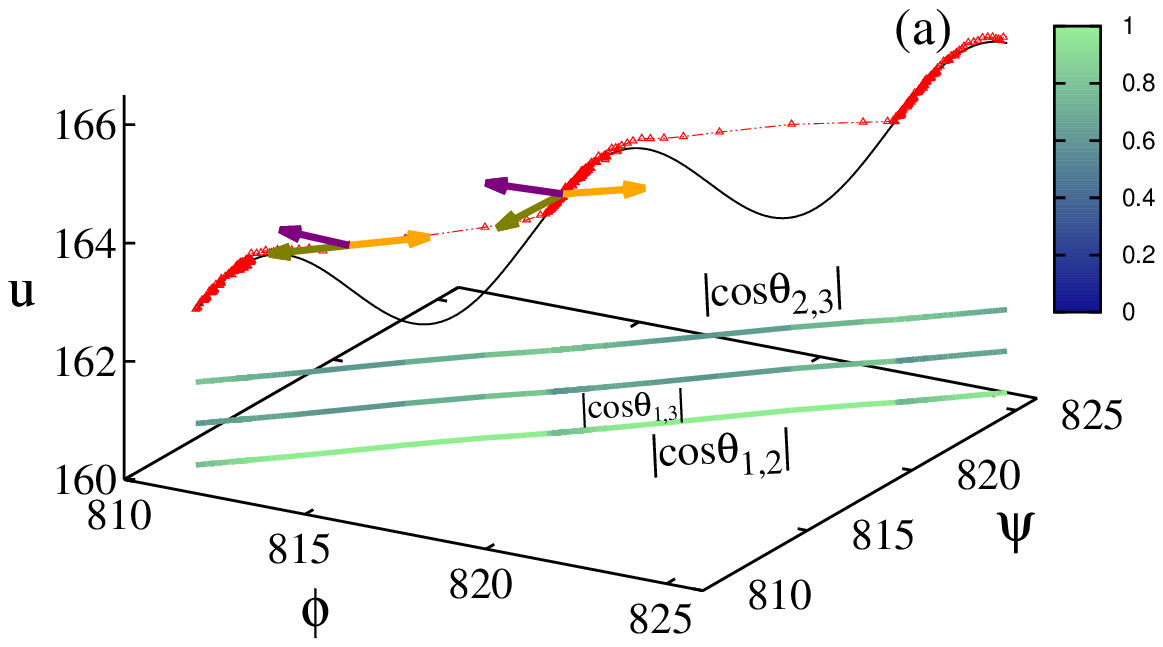}
%\put(-155,92){(b)}
}
\centerline{
\includegraphics[width=1.0\textwidth]{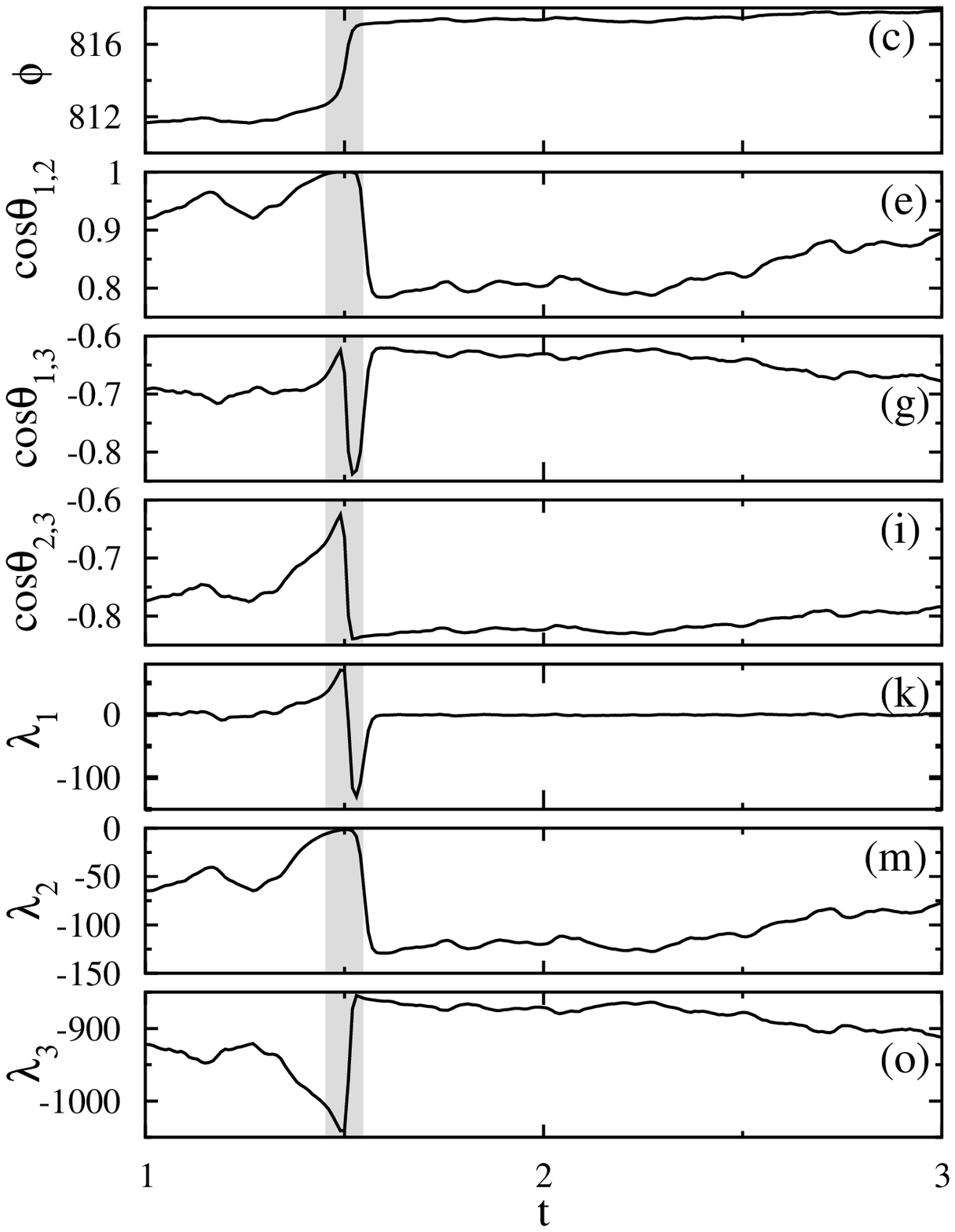}
%\put(-155,92){(b)}
}
\end{minipage}
%
%   \hspace*{0.13cm}
%
\begin{minipage}{0.48\textwidth}
\centerline{
\includegraphics[width=1.0\textwidth]{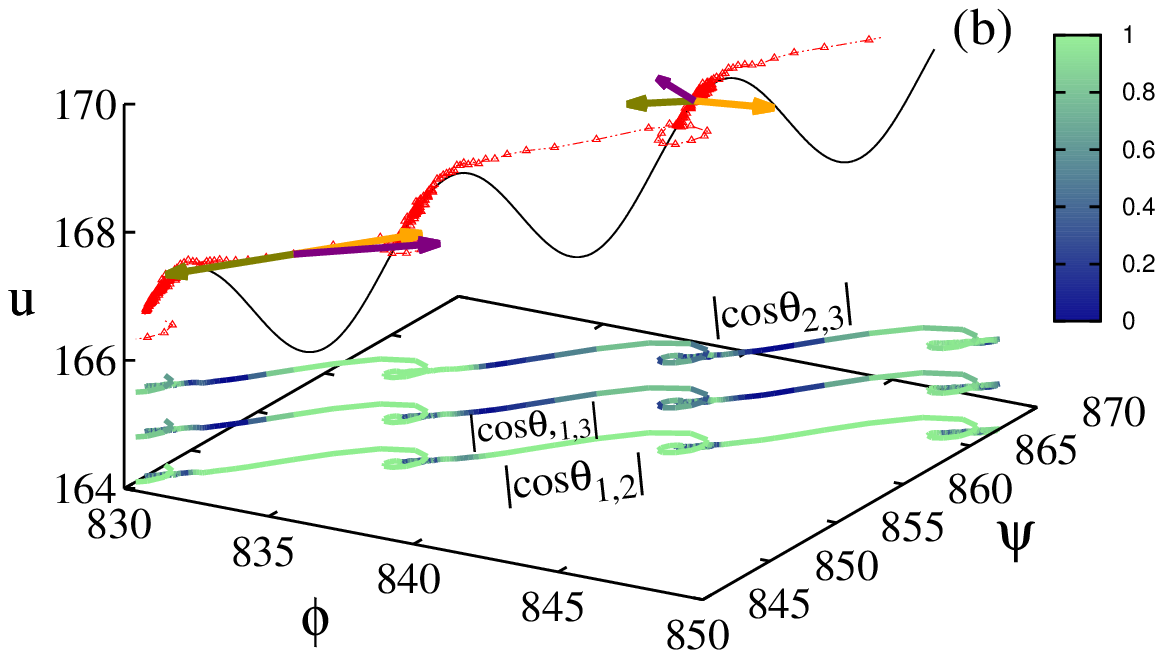}
% \put(-155,92){(c)}
}
\centerline{
\includegraphics[width=1.0\textwidth]{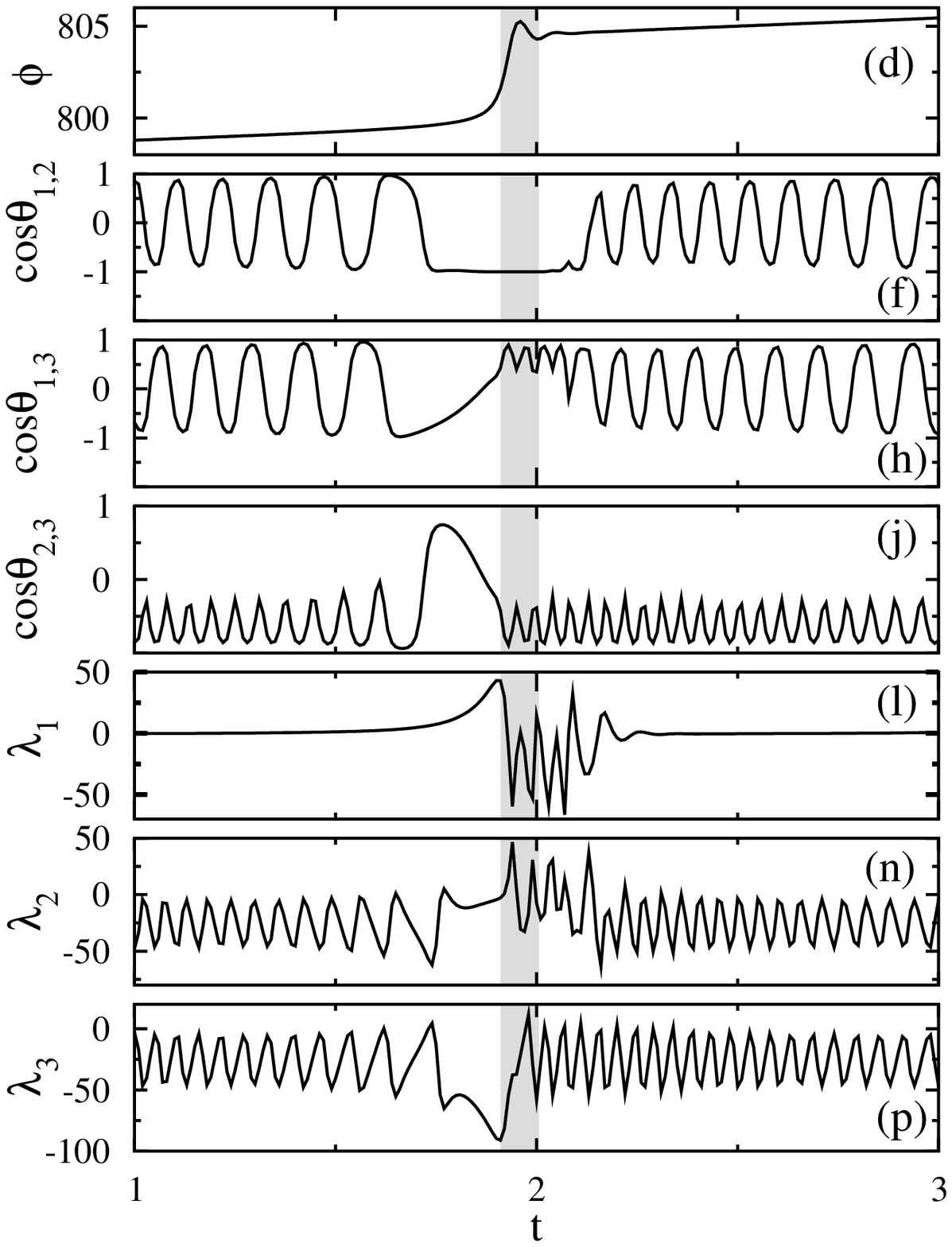}
% \put(-155,92){(c)}
}
\end{minipage}
\caption{\label{figjj}In the model of Josephson junctions, the first and the second vector have tangencies during and prior to critical transitions.
The angle between the first and the third vector decreases as well before transitions occur.
These observations hold for both ranges of $\beta$ although the overall dynamics of covariant Lyapunov vectors is very different for different values of $\beta$.
Left: $\beta = 0.1$, right: $\beta$ = 2.
(a) For $\beta = 0.1$, the Josephson junction is effectively two-dimensional.
(b) For $\beta = 2$, the transitions include spiraling around the nullcline.
In both figures, the green, the orange and the purple vectors represent the first, second and third covariant Lyapunov vectors respectively.
The contour lines are shifted projections of the trajectory onto the $\phi$-$\psi$-plane, showing the absolute value of the cosine of the angles between the covariant Lyapunov vectors.
(c,d) Times series of the fast variable $\phi$.
(e-j) Time series of the cosine of the angles between the covariant Lyapunov vectors.
(k-p) Time series of the finite-time Lyapunov exponents.}
\end{figure*}
The Josephson effect consists of a tunneling current between two superconducting metals \cite{jj, jjb}.
A model that describes this effect in terms of a fast-slow system \cite{jjp} is given by the following equations,

\begin{eqnarray}
\beta \epsilon \dot{\phi} & = & \psi - (1+\beta \epsilon)\phi , \\
\epsilon \dot{\psi} & = & u - \hat{\alpha}^{-1}\phi - \sin\phi, \\
\dot{u} & = & J -\sin\phi + \sqrt{2D} \eta(t).
\label{eq:jj3d}
\end{eqnarray}
%%%%%%%%%%%%%%%%%%%%%%%%%%%%%%%%%%%%%%%%%%%%%%%%%%%%%%%%%%%%%%%%%%%%%%%%%%%%%%%%%%%%%%%%%%%%%%%%%%%%%%%%%%
%%%%%%%%%%%%%%%%%%%%%%%%%%%%%%%%%%%%%%%%%%%%%%%%%%%%%%%%%%%%%%%%%%%%%%%%%%%%%%%%%%%%%%%%%%%%%%%%%%%%%%%%%%
%%%%%%%%%%%%%%%%%%%%%%%%%%%%%%%%%%%%%%%%%%%%%%%%%%%%%%%%%%%%%%%%%%%%%%%%%%%%%%%%%%%%%%%%%%%%%%%%%%%%%%%%%%
%%%%%%%%%%%%%%%%%%%%%%%%%%%%%%%%%%

Assuming $\epsilon \ll 1$, we have a fast-slow system.
However, $\beta$ should be small as well since, for $\beta \gg 1$ (namely $\beta > 10 $), the system does not show fast-slow behavior any more.
Although the system does not have a stable fixed point and transitions happen in the absence of noise as well, adding a stochastic term to the bifurcation parameter $u$ makes the transitions stochastic.
Here, we chose the parameters of the system (other than $\beta$) to be $\epsilon = 0.01, J = 1.5, \hat{\alpha}^{-1} = 0.2$ and $D = 0.2$.

In the limit of $\beta \ll 1$, $\phi$ and $\psi$ are approximately equal and the system becomes effectively two-dimensional.
Examining the system for different values of $\beta$, it can be seen that, as long as $\beta < 0.22$, the system can be reduced to a two-dimensional system [see Fig.~\ref{figjj}(a)].
In this parameter range the covariant Lyapunov vectors and the finite-time Lyapunov exponents show a qualitatively similar behavior as observed in the two-dimensional FitzHugh-Nagumo oscillator.

While the first and the second covariant Lyapunov vector are distinctly tangent during and prior to the critical transition [see Fig.~\ref{figjj}(e)], the third covariant Lyapunov vector, corresponding to the highly negative finite-time Lyapunov exponent [see Fig.~\ref{figjj}(o)], exhibits near-tangencies during the CTs with the other covariant Lyapunov vectors [see Figs.~\ref{figjj}(g,i)], respectively.
The first finite-time Lyapunov exponent shows a spiking behavior during the transition [see Fig.~\ref{figjj}(k)], as we have also observed for the previous models.
The constant increase in the absolute value of the other two finite-time Lyapunov exponents [see Figs.~\ref{figjj}(m,o)] is also comparable to the dynamics of the second finite-time Lyapunov exponent of the FitzHugh-Nagumo and the third finite-time Lyapunov exponent of the Hindmarsh-Rose model.

The second finite-time Lyapunov exponent is close to zero during the transition and the sudden change in the first finite-time Lyapunov exponent will lead to an intersection between the two finite-time Lyapunov 
exponents and a temporary change between the marginal and the stable directions. The temporal change in the stability as discussed before leads to a clear tangency between the first and the second covariant Lyapunov 
vector. 
The third Lyapunov exponent is highly negative. It's finite-time value reaches it's maximum prior to the transition. The decrease in the stability of this direction presents it's self in the decrease in 
the angle between the third and the first as well as the third and the second covariant vector. 
Nonetheless the third finite-time Lyapunov exponent stays substantially smaller than the other two exponents through the transition.
Therefore the third covariant Lyapunov vector remains not tangent to the first two covariant vectors.

For $\beta > 0.22$ the system is no longer two-dimensional.
In this parameter range the values of the Lyapunov exponents are less negative and closer to each other. The second and the third finite-time Lyapunov exponent exhibit oscillations and frequently intersect
resulting the oscillatory changes in the angle between the vectors. Upon further increasing of $\beta$, the first finite-time Lyapunov exponent will also start oscillating and frequently intersecting 
with the two other finite-time exponents. As a result the invariant manifolds frequently  change directions forcing the trajectory to spiral around the nullcline during a transition.
Nonetheless same as our other models, the vectors exhibit clear tangencies during each transition [see Fig.~\ref{figjj}(f)].
%
 
%
%
%%%%%%%%%%%%%%%%%%%%%%%%%%%%%%%%%%%%%%%%%%%%%%%%%%%%%%%%%%%%%%%%
%
\section{Coupled FitzHugh-Nagumo Oscillators}
\label{fhnnet}
%%%%%%%%%%%%%%%%%%%%%%%%%%%%%%%%%%
\begin{figure}[!h]
\begin{subfigure}{}
\includegraphics[width=0.98\linewidth]{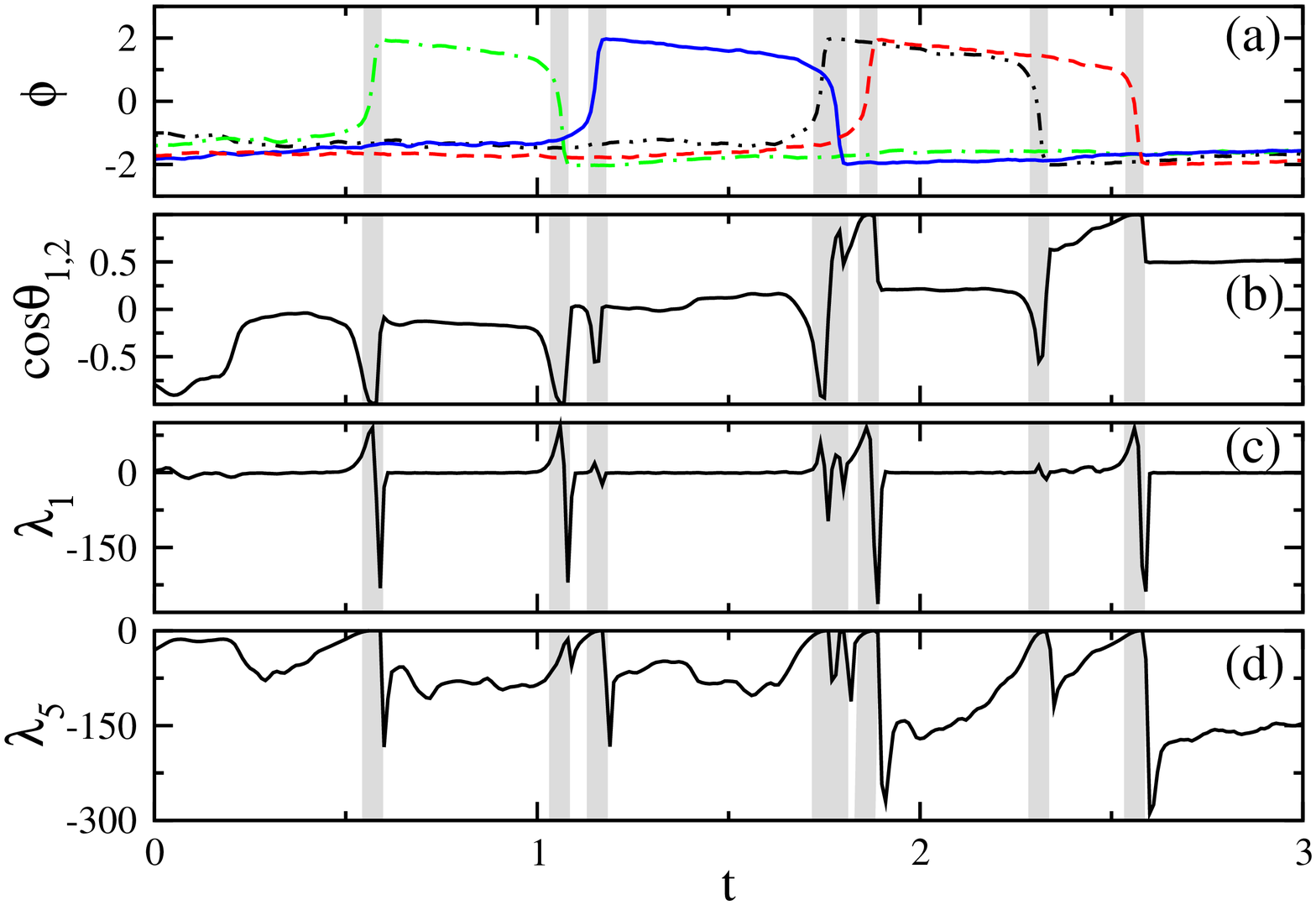}
\end{subfigure}
\vspace{-0.5cm}
\caption{In a network of four coupled FitzHugh-Nagumo oscillators, covariant Lyapunov vectors also align during transitions.
The finite-time Lyapunov exponents exhibit qualitatively the same behavior as observed for a single FitzHugh-Nagumo oscillator $a = 0.4$, $b = 0.3$, $\epsilon = 0.01$ and $D = 0.2$.
(a) Time series of the fast variables of 4 fully connected FitzHugh-Nagumo oscillators going through noise-induced transitions.
(b) Time series of the angle between the first and the second covariant Lyapunov vector.
(c and d) Time series of the first and the fifth finite-time Lyapunov exponents.}
\label{fig:cfhn}
\end{figure}
%%%%%%%%%%%%%%%%%%%%%%%%%%%%%%%%%%%%%%%%
We furthermore studied critical transitions in networks of $N$ coupled FitzHugh-Nagumo oscillators,
\begin{eqnarray}
\epsilon\dot{x_i} & = & x_i - \frac{x_i^3}{3}-y_i + c \sum_{j=1}^{N} K_{ij}(x_j-x_i),  \\
\dot{y_i} & = & x_i + a -by_i + \sqrt{2D}\eta(t). \label{eqfhncoup}
\end{eqnarray}
%
%\textcolor{red}{(one sentence on why study four oscis with noise?)} 
%
% Eq.~(\ref{eqfhncoup}) shows $N$ coupled FitzHugh-Nagumo oscillators.
%
with $K$ being the adjacency matrix and $c$ representing the coupling strength. 
We explored networks of coupled oscillators of different sizes.
However, for simplicity, we merely show four FitzHugh-Nagumo oscillators coupled to each other.

Fig.~\ref{fig:cfhn}(a) shows the time series of the fast variables of four connected oscillators.
As shown in Fig.~\ref{fig:cfhn}(b), the angle between the first and the second covariant Lyapunov vector decreases prior to most of the transitions happening in any of the oscillators.
%
% Tangencies among the covariant Lyapunov vectors are observed during transitions.
%
The angles between higher-order vectors are not shown here, however, they qualitatively exhibit the same behavior.
There are 4 Lyapunov exponents (time-averaged, non finite-time) that correspond in value to the first Lyapunov exponent of the previously studied single FitzHugh-Nagumo oscillator.
These exponents are almost zero or slightly negative and represent the neutral directions.
Additionally, there are four Lyapunov exponents that correspond to the second Lyapunov exponent of the single FitzHugh-Nagumo oscillator.
These exponents are highly negative and represent the contracting directions.
Fig.~\ref{fig:cfhn}(c) shows that the first finite-time Lyapunov exponent is always approximately zero but before and during the first half of the transition, it suddenly increases, followed by a rapid decrease during the second half, accompanied by an overshoot before going back to zero.

Although only the first finite-time Lyapunov exponent has been shown here, all 4 exponents corresponding to the 4 neutral directions exhibit qualitatively the same dynamics.
The 5th finite-time Lyapunov exponent is highly negative as shown in Fig.~\ref{fig:cfhn}(d).
However, it increases prior to and reaches zero during the transition.
Although the last three finite-time Lyapunov exponents have not been displayed here, they exhibit qualitatively the same dynamics during the transition as the 5th exponent.
%
% Furthermore, they become more negative successively.
%
In total, the time series of $\mbox{cos}\,\theta_{ij}$ and of finite-time Lyapunov exponents in all studied fast-slow models indicate that both, orientation and instantaneous growth rates, are sensitive to an upcoming transition.
%
%%%%%%%%%%%%%%%%%%%%%%%%%%%%%%%%%%%%%%%%%%%%%%%%%%%%%%%%%%%
%%%%%%%%%%%%%%%%%%%%%%%%%%%%%%%%%%%%%%%%%%%%%%%%%%%%%%%%%%%
\section{Predicting Critical Transitions}
\label{pred}
%%%%%%%%%%%%%%%%%%%%%%%%
\begin{figure}[t]
\centerline{
\includegraphics[angle=0,width=0.3\textwidth]{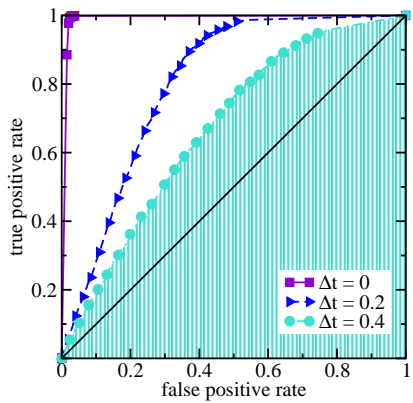}
}
\caption{\label{rocex}
The area under the curve (AUC) is a measure of the effectiveness of predictions.
Increasing the time lag $\Delta t$ decreases the success of predictions.
Different ROC curves for a FitzHugh-Nagumo oscillator with noise strength $D = 0.3$.}
\end{figure}
\begin{figure*}[t]
\begin{minipage}[t]{0.32\textwidth}
\centerline{
\includegraphics[width=1.0\textwidth]{Figure10a.eps}
%\put(-155,92){(a)}
}

\end{minipage}
\hspace*{0.13cm}
\begin{minipage}[t]{0.32\textwidth}
\centerline{
\includegraphics[width=1.0\textwidth]{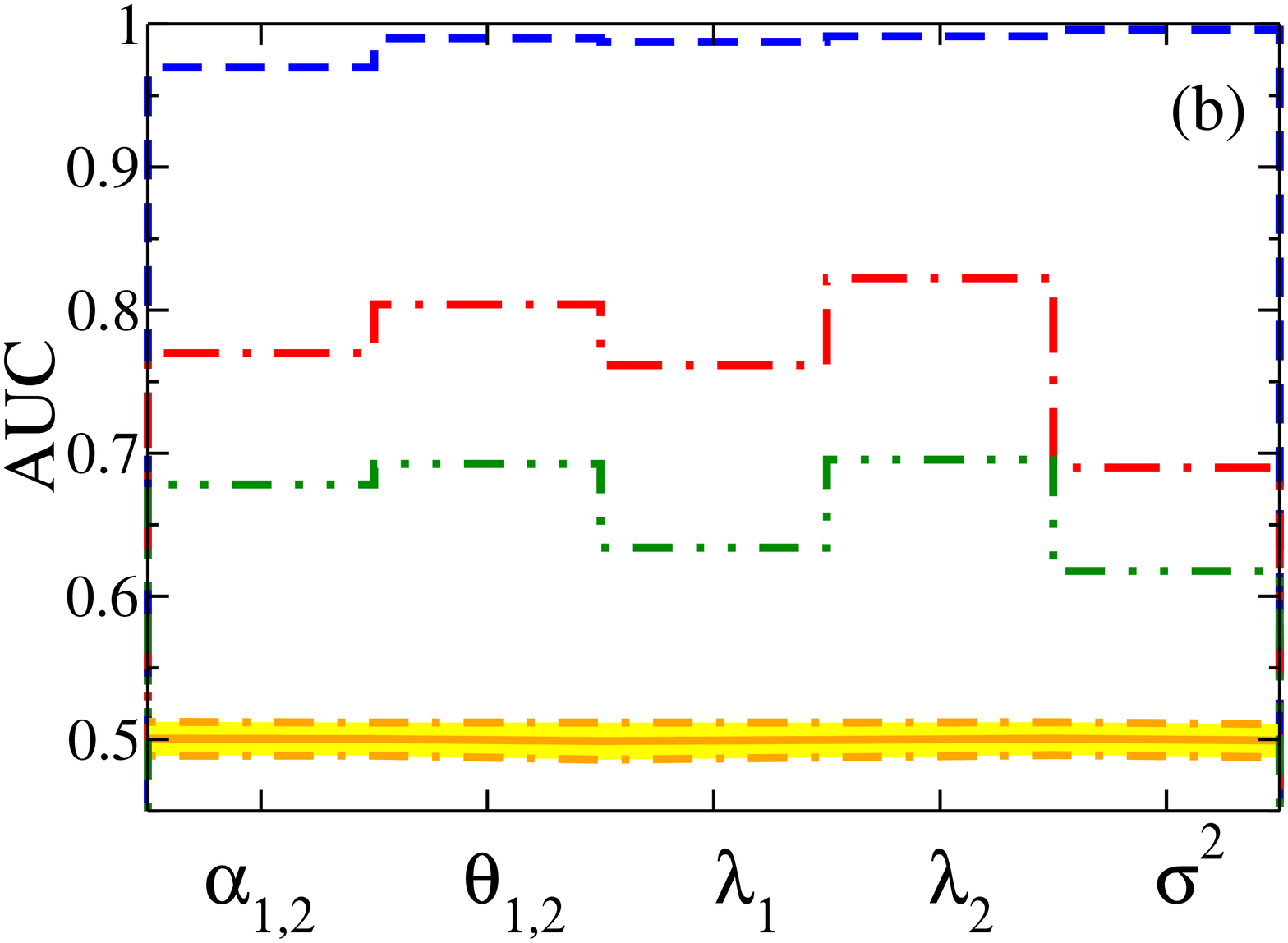}
%\put(-155,92){(b)}
}
\end{minipage}
\hspace*{0.13cm}
\begin{minipage}[t]{0.32\textwidth}
\centerline{
\includegraphics[width=1.0\textwidth]{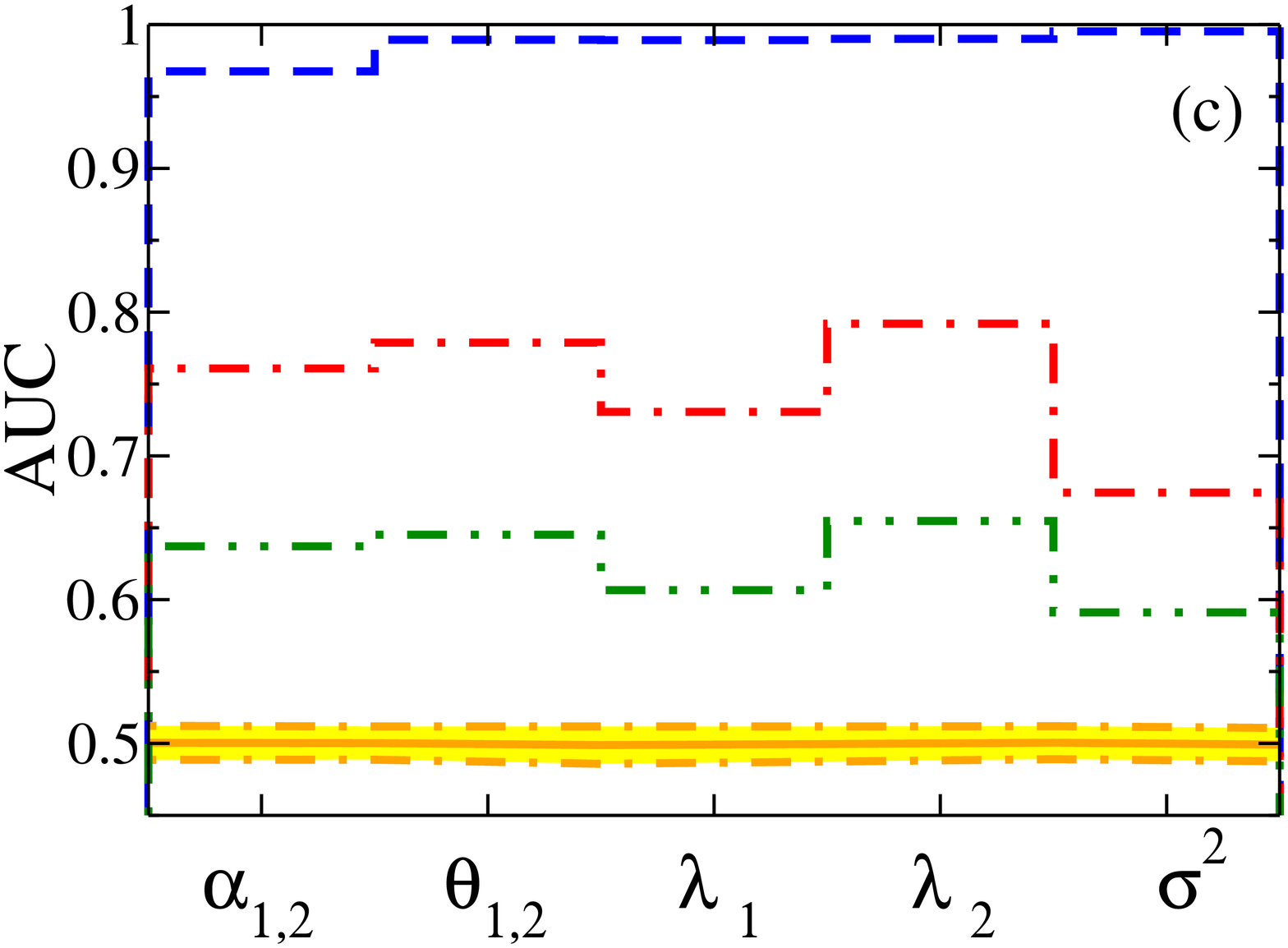}
% \put(-155,92){(c)}
}
\end{minipage}
\caption{\label{fig:fhnauc}
AUCs of the angle between the covariant Lyapunov vectors estimated without iterating to the future, $\alpha_{ij}$, and with Ginelli et al.'s method, $\theta_{ij}$, finite-time Lyapunov exponents, $\lambda$, and the variance, $\sigma$, for different noise strengths, $D$, and different lead times, $\Delta t$, for a single FitzHugh-Nagumo oscillator. 
In the presence of noise, predictors based on covariant Lyapunov vectors can predict CTs better than conventional indicator variables, such as $\sigma^{2}_1$.
The lead time, $\Delta t$, indicates the time lag between the prediction and the event.
The noise strengths being (a) $D = 0$, (b) $D = 0.3$ and (c) $D = 0.6$.
Yellow (gray) shaded regions represent $95$\% confidence intervals estimated from random predictions on the same data sets.}
\end{figure*}
%%%%%%%%%%%%%%%%%%%%%%%%%%%%%%%%%%%%%%%
\begin{figure*}[t]
\begin{minipage}[t]{0.32\textwidth}
\centerline{
\includegraphics[width=1.0\textwidth]{Figure11a.eps}
%\put(-155,92){(a)}
}

\end{minipage}
\hspace*{0.13cm}
\begin{minipage}[t]{0.32\textwidth}
\centerline{
\includegraphics[width=1.0\textwidth]{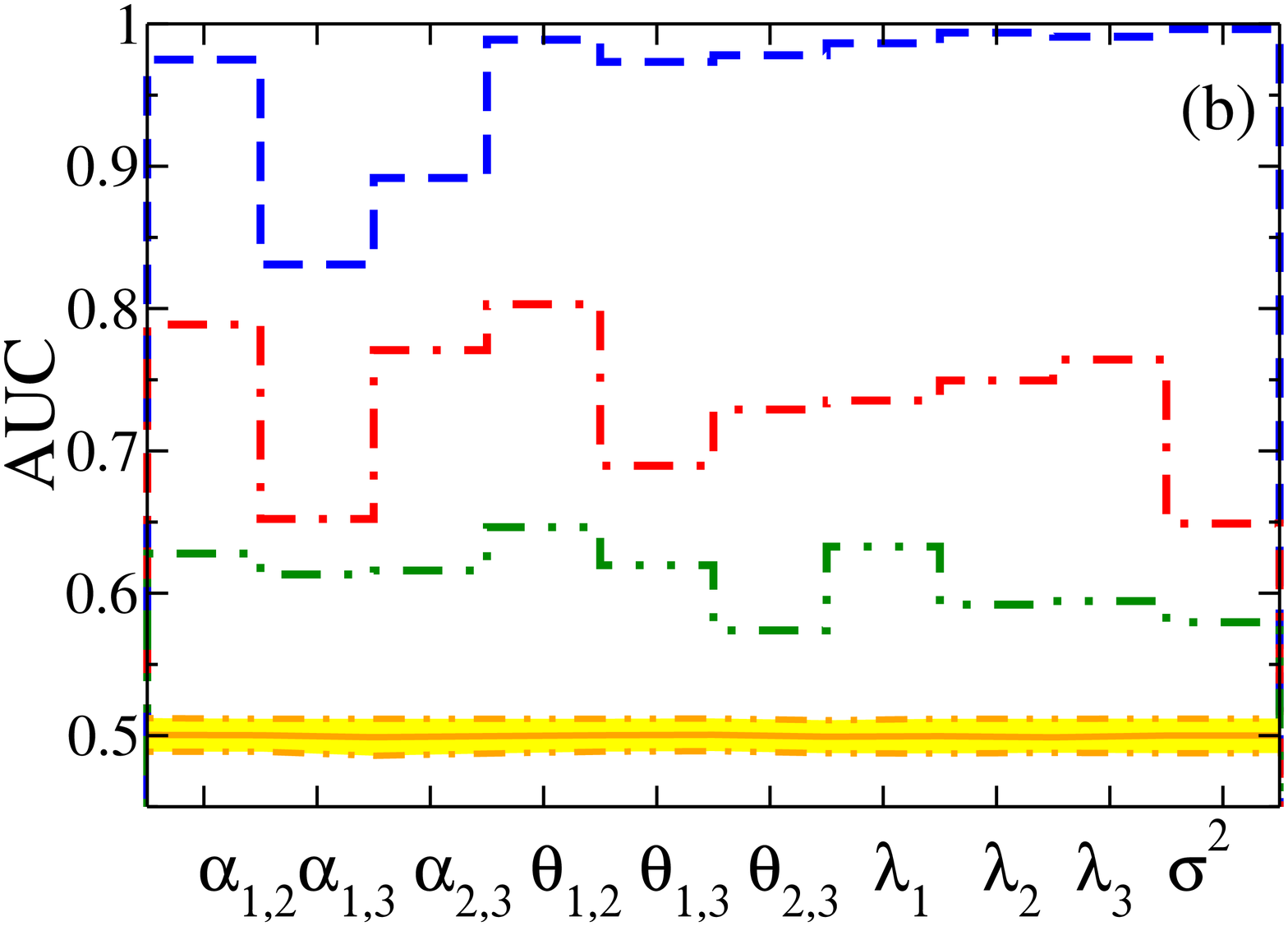}
%\put(-155,92){(b)}
}
\end{minipage}
\hspace*{0.13cm}
\begin{minipage}[t]{0.32\textwidth}
\centerline{
\includegraphics[width=1.0\textwidth]{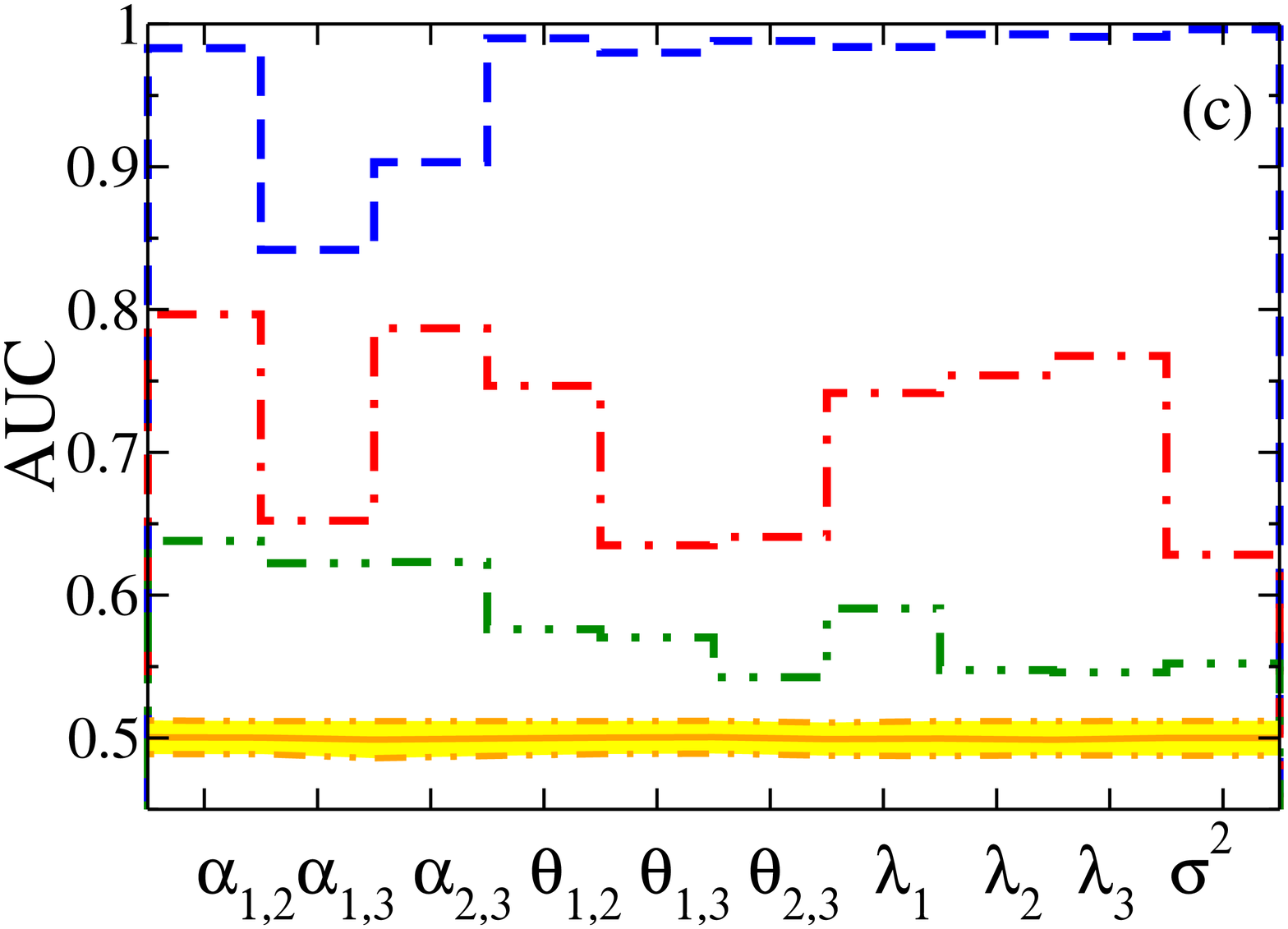}
% \put(-155,92){(c)}
}
\end{minipage}
\caption{\label{fig:jjauc}
AUCs of the angle between the covariant Lyapunov vectors estimated without iterating to the future, $\alpha_{ij}$, and with Ginelli et al.'s method, $\theta_{ij}$, finite-time Lyapunov exponents, $\lambda$, and the variance, $\sigma$, for different noise strengths, $D$, and different lead times, $\Delta t$, for the Josephson junction. 
In the presence of noise, predictors based on covariant Lyapunov vectors can predict CTs better than conventional indicator variables, such as $\sigma^{2}_1$.
The lead time, $\Delta t$, indicates the time lag between prediction and occurrence of the CT.
The noise strengths being (a) $D = 0$, (b) $D = 0.4$ and (c) $D = 0.8$.
Yellow (gray) shaded regions represent $95$\% confidence intervals estimated from random predictions on the same data sets.}
\end{figure*}

%%%%%%%%%%%%%%%%%%%%%%%%%%%%%%%%%%%%%
To demonstrate that the previous observations are not only present in selected short segments of the time series, but are typical and generic instead, we evaluate the existence of links between changes in features of covariant Lyapunov vectors and CTs statistically.
Therefore, we set up prediction experiments in which we use time series of angles between covariant Lyapunov vectors and of finite-time Lyapunov exponents as indicator variables.
Dividing all time series from the simulations described in the previous sections into training and test data sets, containing at least $3\cdot 10^{3}$ transitions each, we evaluate how far both, the angle between the covariant vectors and finite-time Lyapunov exponents, can predict critical transitions.
Similar tests have been used to quantify links in the sense of Granger causality \cite{Granger} between discrete events and continuous variables of dynamical systems \cite{Kim2011,AstridSarah}.
%Similar tests have been used to quantify links in the sense of Granger causality \cite{Granger} between discrete events and continuous variables of dynamical systems \cite{Kim2011,AstridSarah, XiaoZhuCT}.
%
Following these approaches, we apply a simple Bayesian classifier and analyze the success of predictions using Receiver Operating Characteristic curves (ROC curves) \cite{Egan}, which are a common measure for the success of classification algorithms in machine learning and data mining.

In order to identify relevant values of indicator variables, we estimate conditional probability distribution functions on the training data sets.
% %
% For each time series of an indicator variables $\rho(t)$, (i.e., here the time series of angles between the covariant Lyapunov vectorss or finite-time Lyapunov exponents,) we estimate CPDFs $p(\chi(t+\Delta t) = 1| \rho(t))$ with $\chi(t)$ being a binary tracer variable which is unity if a critical transition starts at time $t + \Delta t$ and zero otherwise.
%
% Having estimated CPDFs on training data sets, we can use them to predict CTs in the test data set.
% %
% Therefore, we construct a binary decision variable $\delta \in \left[0,\mbox{max} [p(\chi(t+\Delta t) = 1| \rho(t_{\mbox{test}}))] \right]$.
% %
% This variable is our decision threshold for making predictions, i.e., we issue an alarm for a critical transition occurring at time $t + \Delta t$ whenever we observe the CPDF, $p(\chi(t+\Delta t) = 1| \rho(t_{\mbox{test}}))>\delta$ with $\rho(t_{\mbox{test}})$ being a corresponding observation to $\rho(t)$ in the test data set.
%
% We then count the fraction of correct predictions out of all observed events (true positive rate) and put it into relation to the fraction of false alarms out of all nonevents (false positive rate) for different threshold values $\delta$.
%
% In order to identify the relevant values of indicator variables, we have to es% timate the conditional probability distribution function.
%
In more detail, we use the training data sets, to estimate the conditional probability of an event (a critical transition in this case) happening at time $t_n + \Delta t$ in the future given a certain value of an indicator variable at time $t_n$.

Using these conditional probability distributions, we then predict critical transitions in the test data sets.
Therefore we choose a decision threshold between zero and the maximum of the conditional probability distribution.
%
% After determining the threshold we go through the test data set.
%
For every value the indicator variable assumes in the test data set, we check if the conditional probability associated with this value is above the threshold.
If this condition is fulfilled, we predict a critical transition to occur $\Delta t$ time steps in the future.
We then count the fraction of correct predictions out of all observed events (true positive rate) and put it into relation to the fraction of false alarms out of all nonevents (false positive rate) for different threshold values.
These two rates generate a point on the ROC-curve.
For each value of the decision threshold we obtain a single point on the ROC curve [see Fig.~\ref{rocex}].
Varying the decision threshold from zero to the maximum of the conditional probability distribution, we obtain the complete ROC-curve.

For random predictions, the rate of correct predictions and the rate of false alarms will be approximately the same, resulting a diagonal (black-line in Fig.~\ref{rocex}).
The more effective and non-random the predictions are, the closer the ROC curve will get to the upper left corner.
A measure that summarizes each ROC curve and allows for comparison of several indicator variables is the area under the curve (AUC), which ranges from zero to unity.
We compute AUCs for the angle between the covariant Lyapunov vectors, for finite-time Lyapunov exponents and, additionally, for the sliding window variance, which is a well known indicator for critical slowing down \cite{Scheffer}.
In order to test the robustness of the precursors, we additionally compute AUCs for different lead times (times between issuing an alarm and the occurrence of the transition) $\Delta t$, and different noise strengths $D$, in the respective dynamical models.
In order to estimate $95$\% confidence intervals for AUCs, we additionally compute $100$~AUCs, generated by making random predictions (yellow areas in Fig.~\ref{fig:fhnauc} and Fig.~\ref{fig:jjauc}) for each test data set.
As candidates for indicator variables we use the following quantities:
\vspace{-0.2cm}
\begin{itemize}
\item[$\alpha_{ij}$] the angle between the covariant Lyapunov vectors estimated without iterating to the future (see Sec.~\ref{newmeth}), 
\vspace{-0.2cm}
\item[$\theta_{ij}$] the angle between the covariant Lyapunov vectors calculated using Ginelli et al.'s method (see Sec.~\ref{compmeth})
\item[$\lambda_i$] finite-time Lyapunov exponents (see Sec.~\ref{ftles}) and
\item[$\sigma$] and the sliding-window-estimate of the variance.
\end{itemize}
Note that the indices $i$ and $ij$ indicate the order of the respective Lyapunov vectors and exponents and vary according to the dimension of the system under study.
In order to calculate covariant vectors and their approximations, we used a sampling interval of $dt = 0.001$.
We orthogonalized perturbations every 10 time steps and simulated in total between $7\cdot 10^{5}$ and $6\cdot 10^{6}$ time steps in order to record at least $6\cdot 10^{3}$ transitions within each data set.
Half of each data set is used for training, i.e., estimating conditional probabilities, and the other half for predicting transitions and computing ROCs and AUCs.
In order to obtain estimates of the sliding-window variance which is a common indicator for critical transitions, we used a sliding window average of $10$ steps.

Fig.~\ref{fig:fhnauc} and Fig.~\ref{fig:jjauc} show the results for predictions of transitions in a single FitzHugh-Nagumo oscillator and in a model for Josephson junctions for different noise strengths (increasing from left to right), and different lead times, $\Delta t$.
As the AUCs in Fig.~\ref{fig:fhnauc} indicate, all predictions are far better than random predictions (ROC on the diagonal, $\text{AUC}=1/2$).
For all indicator variables and models tested we found that AUCs obtained from the indicator variables we used are outside the 95\% confidence bands estimated by making random predictions within the test data set.
Consequently we can conclude that their exist a Granger causal links between the dynamics of the angles between covariant Lyapunov vectors (i.e., the occurrence of tangencies) and the occurrence of critical transitions.

For small lead times, $\Delta t$, and without noise ($D=0$), we observe predictions that are very close to the optimal value, $\text{AUC}\approx1$, as it is expectable for a system without noise.
Increasing the lead time and the noise strength decreases the prediction's success.
For larger lead times and increased noise strength angles between covariant Lyapunov vectors, their approximations and finite time Lyapunov exponents lead to better predictions than using the sliding window estimate of the variance.
This seems surprising since the variance is typically considered to be a very robust indicator for CTs and it indicates the potential of $\alpha_{1,2}$ and $\theta_{1,2}$ and finite time Lyapunov exponents as indicator variables.
It is also surprising, that the results for the Josephson junctions indicate that for large noise strength the approximated angles $\alpha_{i,j}$ leads to better predictions than the angles computed through Ginelli's method $\theta_{i,j}$. 
One reason for this could be the fact that the approximated angles are computed using only information from the present and the close past and are therefore more sensitive to the onset of a transition than $\theta_{i,j}$.
%
%%%%%%%%%%%%%%%%%%%%%%%%%%%%%%%%%%%%%%%%%%%%%%%%%%%%%%%%%%%%%%%%%%%%%%%%%%
%%%%%%%%%%%%%%%%%%%%%%%%%%%%%%%%%%%%%%%%%%%%%%%%%%%%%%%%%%%%%%%%%%%%%%%%%%%
\section{Conclusions}
\label{conclusions}
Extreme events and critical transitions have been discussed from a dynamical systems perspective in the past \cite{albeverio2006extreme,aytacc2015laws,lucarini2014towards,lapeyre2002characterization,beims2016alignment,pazo2010characteristic,cencini2013lyapunov}.
In this contribution we investigated the time resolved behavior of covariant Lyapunov vectors with respect to the onset of critical transitions, as modeled by fast-slow systems.

We verified that the alignment of covariant Lyapunov vectors can be linked to critical transitions in a Granger causal sense.
Verification was done by carrying out a set of prediction experiments which consist of identifying indicatory behavior of time series derived from covariant Lyapunov vectors in training data sets, predicting critical transitions occurring in test data sets and evaluating the prediction success using common measures of forecast verification. 
The testing was necessary, since in systems with increased noise strength some transitions could potentially occur without previous alignment or some alignments might occur without a following critical transition.
For all prediction experiments we found that the angle between covariant Lyapunov vectors was able to predict the occurrence of critical transitions significantly better than chance. 
In order to verify the existence of a Granger causal link, it is sufficient to verify that the predictions based on the alignment are better than random predictions. 
This condition is fulfilled for all indicator variables related to (covariant) Lyapunov vectors. 
That is, we found a Granger causal link between the alignment of covariant Lyapunov vectors prior to and during critical transitions and the occurrence of critical transitions. 
Additionally we found another Granger causal link between the specific dynamics in the time series of finite-time Lyapunov exponents prior and during CTs and the occurrence of critical transitions.
For systems with increased noise strength and predictions with longer lead time the angle between covariant Lyapunov vectors predict critical transitions even better than common indicator variables for critical slowing down, such as the sliding-window estimate of the variance.

Summarizing observations and all prediction experiments, we found a generic behavior for all systems studied: covariant Lyapunov vectors align during CTs, which corresponds to the observations in \cite{beims2016alignment}.
The alignment can also be described in terms of the angle between stable and marginal manifold(s), i.e., tangencies or near tangencies can be understood as deviations from hyperbolicity \cite{newhouse1979abundance,gonchenko2002newhouse,yang2011newhouse, HongLiuYang2009, Morris2013}.
We argued that a rapid change in the stability of the marginal manifold is generic to critical transitions. 
Further more we showed that rapid changes in the stability are accompanied by a change in the order of the finite-time Lyapunov exponents that results in tangencies between covariant Lyapunov vectors.
The alignment can also be understood as an alignment of stable and marginal manifold.

Considering the observed phenomena with respect to known approaches to critical transitions, studying covariant Lyapunov vectors allows us to obtain additional insight to the  mechanism typically referred to as critical slowing down.
Critical slowing down describes the fact that a system approaching a bifurcation point looses its resilience to external perturbations, i.e., the rate of recovering from external perturbations slows down.
In this contribution we discover that it is not only the increase in the growth rates of perturbations, but also constraints on possible growth directions, that cause critical slowing down.
The alignment of covariant Lyapunov vectors acts as a temporal reduction of the dimension of the tangent space in which perturbations can grow. 
During and close to the transitions this tangent space is effectively one-dimensional, allowing only perturbation growth in the direction of the trajectory, i.e., in the direction of the transition. 
In other words, shortly before the transition, while the vectors tend to align, any perturbation in any direction will grow such that it triggers the transition. 
During the transition any perturbation will grow such that it is contributing to the transition.
In fact, the alignment of both possible directions of perturbation growth, previous to and during the CTs, indicates that the dimension of the tangent space is reduced during the transition, allowing only one possible change of the trajectory: towards the next (meta) stable state.

%
%Complementing the observations on two very similar deterministic $3$-dimensional% models in \cite{beims2016alignment}, we studied several deterministic and stoch%astic models exhibiting CTs.
%fme
%Furthermore, we extended our studies to all angles between covariant Lyapunov vectors and finite-time Lyapunov exponents and we% also investigated.
%

%
The results of the prediction experiments demonstrate that giving alarms for critical transitions based on alignments of covariant Lyapunov vectors can predict the occurrence of critical transitions better than or equally as well as common indicator variables associated with critical slowing down such as the sliding-window variance.  
Furthermore, we proposed and tested a method for estimating approximations of covariant Lyapunov vectors which allows to obtain an indicator variable (the angle between vectors) without knowing the far future of the system. 
The results of the prediction experiments indicate that this approximated indicator variable is suitable to be applied in predictive settings.
From a practical perspective one can also ask, in how far these observations can be used in order to predict critical transitions.
When tested against the sliding-window variance (a common indicator variable for critical slowing down), we found that in the presence of noise, using the angle between the vectors or its approximation results in better predictions.
This effect becomes especially visible for increased lead times, i.e., larger time lags between prediction and occurrence of the event.
Also, finite-time Lyapunov exponents can in some cases lead to improved predictions compared to the sliding-window variance.
Although covariant Lyapunov vectors and finite-time Lyapunov exponents have been computed from models in this contribution, there are approaches to estimate them from times series \cite{KantzRadons,wolf1985determining,sano1985measurement,eckmann1986liapunov,darbyshire1996robust,parlitz1992identification}.
Therefore, it would be interesting to study in future contributions how robust the results concerning the quality of the predictions are if covariant Lyapunov vectors and finite-time Lyapunov exponents are estimated from data records.
%
%
%\bibliography{sharafhaller}
%

%%%%%%%%%%%%%%%%%%%%
\end{document}